\documentclass[apj]{emulateapj}

\slugcomment{Accepted for publication in The Astrophysical Journal}

\shorttitle{Subaru Super Deep Field with AO I}
\shortauthors{Minowa et al.}

\begin{document}


\title{Subaru Super Deep Field with Adaptive Optics I.\\ Observations and First Implications\footnote{Based on the data
corrected at the Subaru Telescope, which is operated by the National
Astronomical Observatory of Japan.}}


\author{Yosuke Minowa, Naoto Kobayashi, Yuzuru Yoshii}
\affil{Institute of Astronomy, School of Science, University of Tokyo,
2-21-1 Osawa, Mitaka, Tokyo 181-0015, Japan.}
\email{minoways@ioa.s.u-tokyo.ac.jp}
\author{Tomonori Totani, Toshinori Maihara, Fumihide Iwamuro}
\affil{Department of Astronomy, Kyoto University, Kitashirakawa, Kyoto
606-8502, Japan}
\author{Hideki Takami, Naruhisa Takato, Yutaka Hayano, Hiroshi Terada,
Shin Oya}
\affil{Subaru Telescope, National Astronomical Observatory of Japan, 650
North A'ohoku Place, Hilo, HI 96720, USA}
\author{Masanori Iye}
\affil{Optical and Infrared Astronomy Division, National Astronomical
Observatory of Japan, 2-21-1 Osawa, Mitaka, Tokyo 181-8588, Japan}
\and
\author{Alan T. Tokunaga}
\affil{Institute for Astronomy, University of Hawaii, 2680 Woodlawn Dr.,
Honolulu, HI 96822, USA}

\begin{abstract}
We present a deep $K^{\prime}$-band (2.12$\mu$m) imaging of 1\arcmin\
 $\times$ 1\arcmin\ Subaru Super Deep Field (SSDF) taken with the Subaru
 adaptive optics (AO) system. Total integration time of 26.8 hours
 results in the limiting magnitude of $K^{\prime} \sim 24.7$ (5$\sigma$,
 0\farcs2 aperture) for point sources and $K^{\prime} \sim 23.5$
 (5$\sigma$, 0\farcs6 aperture) for galaxies, which is the deepest limit
 ever achieved in the $K^{\prime}$ band. The average stellar FWHM of the
 co-added image is 0\farcs18.  Based on the photometric measurements of
 detected galaxies, we obtained the differential galaxy number counts,
 for the first time, down to $K^{\prime} \sim 25$, which is more than
 0.5 mag deeper than the previous data. We found that the number count
 slope $d\log N/dm$ is about 0.15 at $22 < K^{\prime} < 25$, which
 is flatter than the previous data. Therefore, detected galaxies in the
 SSDF have only negligible contribution to the near-infrared
 extragalactic background light (EBL), and the discrepancy claimed so
 far between the diffuse EBL measurements and the estimated EBL from
 galaxy count integration has become more serious . The size
 distribution of detected galaxies was obtained down to the area size of
 less than 0.1 arcsec$^2$, which is less than a half of the previous
 data in the $K^{\prime}$ band. We compared the observed size-magnitude
 relation with a simple pure luminosity evolution model allowing for
 intrinsic size evolution, and found that a model with no size evolution
 gives the best fit to the data. It implies that the surface brightness
 of galaxies at high redshift is not much different from that expected
 from the size-luminosity relation of present-day galaxies.
\end{abstract}



\section{Introduction}
The process of galaxy formation and evolution is one of the most
important unsolved problems in astrophysics. Deep imaging of blank field
is a vital method for studying the properties of galaxies in the distant
universe \citep{yos88}. The Hubble Deep Field (HDF) taken by the Hubble
Space Telescope (HST) had revealed deep images of the universe at
optical wavelengths \citep{wil96,wil00,gdr00}, and provided us with
valuable information of the distant universe. However, for high-redshift
galaxies at $z>1$, the rest-frame optical light, that exhibits the
fundamental structure of stellar component in galaxies, shifts into the
infrared (e.g. \citealt{she03}). Therefore, the near infrared (NIR) deep
imaging becomes important to study the formation and evolution of
galaxies at high-redshifts. Moreover, the uncertainties due to the
evolution of galaxies and extinction by interstellar dust are less
significant in the longer wavelength. Thus, deep imaging in the $K$
(2.2$\mu$m) band, that is the longest wavelength at which the
high-sensitive observations can be carried out from the ground, is
essential to study the fundamental properties of high-redshift galaxies.

A number of $K$-band imaging surveys were carried out using ground-based
large telescopes with different spatial coverages and limiting
magnitudes (e.g.
\citealt{gdr93,gdr96,glz94,mcl95,djr95,hua97,mou97,szo98,min98a,ber98,src99,vai00,mar01,mai01,bak03,crh03,lab03}).
Among these surveys, the deep $K^{\prime}$ (2.12$\mu$m) imaging of the
Subaru Deep Field \citep[SDF,][]{mai01} using the Subaru/CISCO
\citep{mot02} achieved a limiting magnitude of $K^{\prime} \sim 23.5$
(5$\sigma$) for point sources with integration time of 9.7
hours. \citet{tot01a} studied the galaxy number counts in the SDF and
suggested that a number evolution may occur or a new population may
emerge in the faint end, while the number count at brighter magnitude
range is consistent with the pure luminosity evolution (PLE) without
number evolution. \citet{tot01b} derived the contributed flux of
galaxies to the extragalactic background light (EBL) in the $K$ band
using the SDF number count data. They concluded, from the observed flat
slope of differential galaxy number count ($d\log N/dm \sim 0.23$) in
the faint end and the theoretical estimate of the number of missed
galaxies in the SDF survey, that more than 90\% of the galaxies
contributed to the EBL in the $K$ band has already been resolved into
discrete sources of galaxies. However, the EBL flux derived in this way
accounts only for less than a half of the EBL flux measured in the same
$K$ band in the form of diffuse emission
\citep{gor00,wri01,mat01,cam01}, indicating a problem of missing $K$
light in the universe, which cannot be explained by normal
galaxies. Therefore, even deeper imaging is required to investigate the
galaxy population in the faint end and the origin of missing $K$ light
in the EBL. Recently, a deep imaging of the Hubble Deep Field South
(HDF-S) was carried out in the $K_s$ band (2.16 $\mu$m) using the
VLT/ISAAC \citep[FIRES,][]{lab03} and reached a limiting magnitude of
23.8 mag (5$\sigma$) with integration time of 35.6 hours. However,
because unrealistically long integration time is required to increase
the sensitivity further from this level, the sensitivity of deep imaging
observations under usual seeing condition has almost reached the
attainable limit.

To push the limit of deep NIR imaging, we initiated a new deep imaging
program in the $K^{\prime}$ band using the Subaru adaptive optics (AO)
system.  AO compensates the disturbed wavefront by earth's atmosphere
and provides nearly diffraction-limited spatial resolution. Because the
flux in the diffraction-limited core largely increases, it is expected
to improve the sensitivity of detecting faint objects with AO.  Although
the sensitivity gain with AO is known to be small for extended objects
such as galaxies, we can improve the sensitivity of detecting
high-redshift galaxies because they are expected to be compact according
to the prediction of hierarchical cold dark matter (CDM) model. Studies
of high resolution deep imaging by the HST/NICMOS have also shown that
faint galaxies at high redshifts are quite compact in the $H$
(1.6$\mu$m) band \citep{yan98}.

First objective of this program is to investigate the galaxy population
in the unprecedented faint end ($K^{\prime} > 24$) with high detection
completeness. This may reveal a significant faint galaxy population that
contributes to the EBL in the $K^{\prime}$ band. Second objective is to study
the morphology of high-redshift galaxies at {\it rest-frame optical
wavelengths} with high-resolution AO image. Formation of the Hubble
sequence is thought to have occurred at $1<z<2$
\citep{kaj01}. High-resolution deep $K^{\prime}$ imaging of galaxies in this
redshift range may clearly reveal the process of morphology formation of
galaxies.

 In this paper, we report the first results of our deep $K^{\prime}$ imaging
program with the Subaru AO system.  The layout of the paper is as
follows. The field selection and observational strategy are summarized
in \S 2. The procedure of data reduction is described in \S 3. The
procedure of source detection and photometry is described in \S 4. The
analysis of reduced images and detected source counts is given in \S
5. The quality of our AO deep $K^{\prime}$ imaging is examined in \S 6. The
results of the galaxy number counts, the contributed flux of galaxies to
the EBL, and the size distribution of detected galaxies are discussed in
\S 7. Finally, the summary is given in \S 8.  Throughout this paper, we
adopt the magnitude system where Vega is 0.0 mag, and the cosmological
parameters of $\Omega_M$=0.3, $\Omega_{\Lambda}$=0.7, and $H_0$=70
Km/s/Mpc.

\section{Observations}
\begin{figure}
\plotone{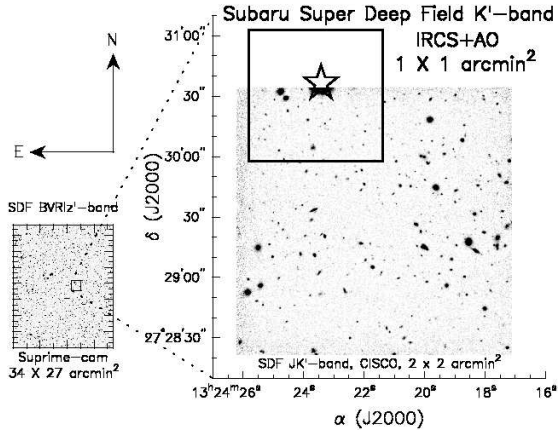}
\caption{The location of the Subaru Super Deep Field (SSDF) relative to
the Subaru Deep Field (SDF). Intensive observations have been carried
out in the SDF using the Subaru Suprime-Cam ($BVRi^{\prime}z^{\prime}$)
and the CISCO ($JK^{\prime}$). The stellate symbol in the SSDF shows the
location of the AO guide star ($R$ $\sim$ 12). \label{field}}
\end{figure}

Our observed field is a part of the well-studied deep field called
``Subaru Deep Field (SDF)'', which is a blank sky region near the north
galactic pole \citep{mai01}. This field has been observed extensively by
the Subaru Telescope \citep{iye04} as an observatory project with a wide
wavelength coverage from optical to NIR
\citep{mai01,kas04}. $JK^{\prime}$-band images for a 2\arcmin\ $\times$
2\arcmin\ area of the SDF were obtained with the
Subaru/CISCO \citep{mot02} and $BVRi^{\prime}z^{\prime}$-band images for
a 30\arcmin\ $\times$ 37\arcmin\ area, which includes the CISCO field,
were obtained by the Subaru/Suprime-cam \citep{miy02}. The SDF was
originally chosen such that a reasonably bright star is located in the
field as a wavefront guide star for AO observation. We placed this
bright ($R \sim 12$) star near the center of our 1\arcmin\ $\times$
1\arcmin\ AO field. The center coordinates of our AO field was set to be
$\alpha$=13$^h$24$^m$23\fs6 and $\delta$=+27\degr30\arcmin30\farcs4
(J2000). We call this field ``Subaru Super Deep Field (SSDF)''. The
location of the SSDF relative to the SDF is shown in Figure
\ref{field}. Because very deep $BVRi^{\prime}z^{\prime}$ images of the
SDF have already been obtained, it is possible to determine the
photometric redshift of detected galaxies in the SSDF. Seeing-limited
infrared $J$ and $K^{\prime}$ images of the SDF have also been taken by
\citet{mai01} and a part of this field is overlapped with our AO field
(see Figure \ref{field}). Thus, the quality of our AO deep images can be
directly compared to that of conventional seeing-limited deep images
(see \S 6).

Our observations were carried out using the Subaru AO system
\citep{tak04} and the IRCS \citep[Infra-Red Camera and
Spectrograph,][]{tok98, kob00} both mounted on the Cassegrain focus of
the Subaru Telescope. The IRCS imager is equipped with a Raytheon
Aladdin III 1024$\times$1024 InSb array, offering two pixel scales of 23
and 58 mas for imaging in a wavelength range of 0.9$-$5.5 $\mu$m. We
used the pixel scale of 58 mas to cover a wider field of view with the
$K^{\prime}$ filter (1.96$-$2.30 $\mu$m, $F_{\lambda} = 4.66 \times
10^{-10}$ Wm$^{-2}\mu$m$^{-1}$, \citealt{tok02}). The Subaru AO system
uses a curvature sensor with 36 control elements, which provides a
stellar image of Strehl ratio (the ratio of the observed star peak to
the peak value of a perfect telescope diffraction pattern) of
$\sim$\,0.28 and full-width at half maximum (FWHM) of $\sim$\,0\farcs07
in the $K$ band with an $R \sim 12$ guide star under best observing
condition. The improved image quality is expected to increase the
detection sensitivity to a point source by $\sim$ 1 mag. However,
because AO correction performance degrades with increasing distance from
a guide star (see \S 6.3 for details), the Strehl ratio at the area away
from a guide star is lower than the best value. Moreover, since the
Strehl ratio should vary with time depending on the observational
condition, the resultant Strehl ratio for long exposure time is averaged
and lower than the best value. Similarly, the FWHM for long-time
exposure image could be broader than the best value at the area away
from a guide star. For the present data, the achieved Strehl ratio and
the FWHM, which were measured from a point-like source located at 24
arcsec away from the guide star, are 0.1\footnote{Since the pixel scale
of 58 mas is not small enough to sample the small size of point spread
function, the estimated Strehl ratio may not be accurate.} and 0\farcs18
on average, respectively. Because the galaxies in our field of view are
distributed within about 40 arcsec away from the guide star, our
measured Strehl ratio and FWHM are typical for our observation.

\begin{figure*}
\begin{center}
\includegraphics[width=0.8\textwidth,clip]{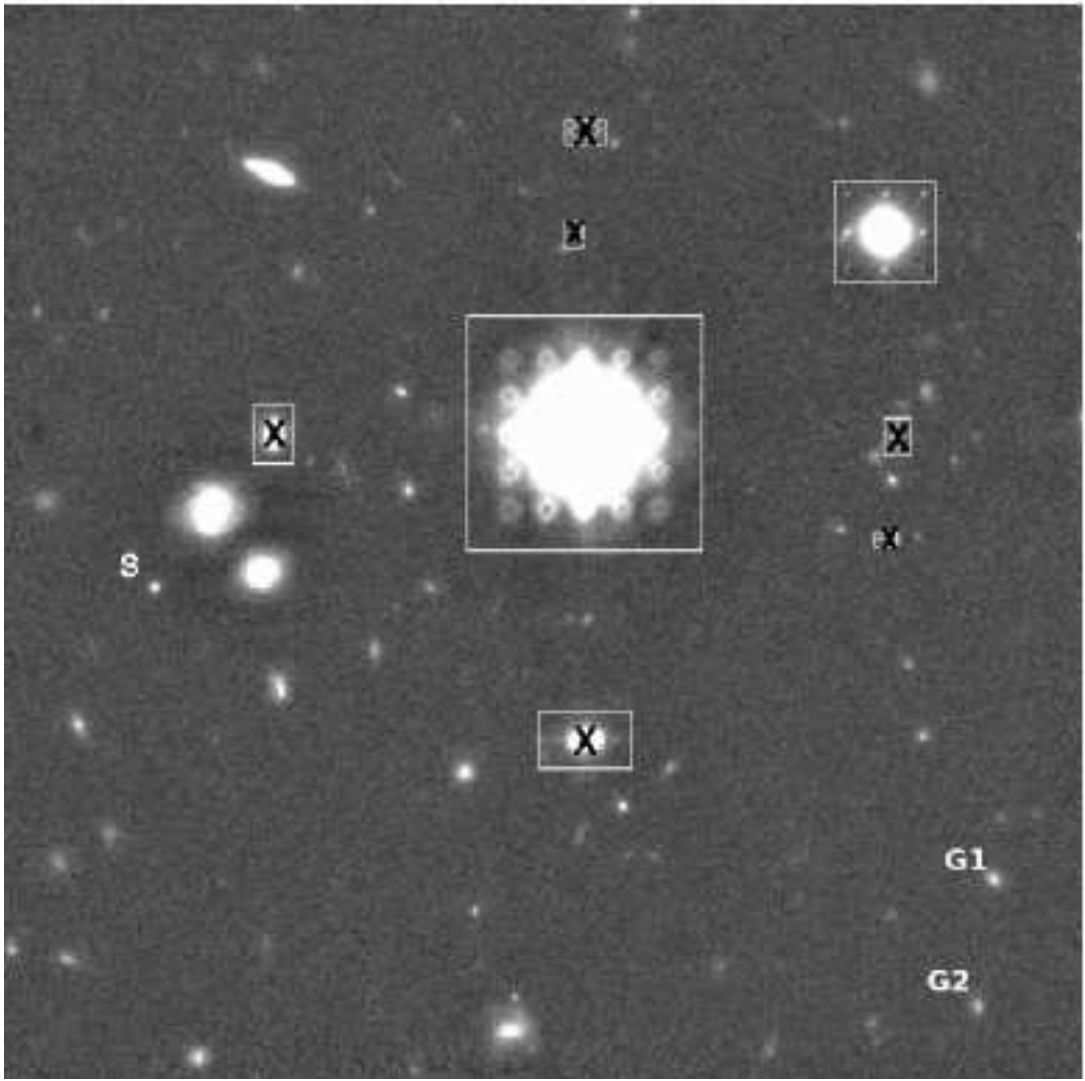}
\end{center}
\caption{IRCS+AO $K^{\prime}$-band image of the Subaru Super Deep Field
(SSDF) with the total integration time of 26.8 hrs. The field size is
1\arcmin\ $\times$ 1\arcmin\ with the pixel scale of 58 mas. The stellar
FWHM was measured to be 0\farcs18 for the point-like source (indicated
by ``S'' mark). The bright star at the center of the image was used as
the AO guide star. Ghosts of bright stars due to the IRCS internal
optics (beam splitter and compensator) are indicated by cross marks. The
grid patterns around the guide star and other bright objects are the
residual image of previous frames due to saturation of the
detector. Because of these mock objects, some areas near the bright
stars and their ghosts (boxed areas) were excluded from the detection
area. G1 is a disk galaxy with an effective radius of 0\farcs25, that is
the average size of galaxies in the SSDF image (see \S6.2). G2 is a
galaxy whose radial profile and model fitting are shown in Figure
\ref{radfit}. \label{ssdf_image}}
\end{figure*}

We carried out our observations for a total of nine nights with almost
photometric condition. The observing log is given in Table
\ref{log}. During the observations, we repeated the set of nine-position
dithering with a 3$\times$3 grid of 2\farcs0 separation. Individual
exposure time of 90 or 120 sec was adopted such that the sky background
does not saturate the detector well. Since a few bright stars reached
saturation even with this short exposure time, residual images
following the dithering pattern appeared around the stars due to the
memory effect of the InSb array (see grid patterns around bright stars
in Figure \ref{ssdf_image}).
 
\begin{deluxetable}{ccc}
\tabletypesize{\scriptsize}
\tablecaption{Summary of $K^{\prime}$-band observations. \label{log}}
\tablewidth{0pt}
\tablehead{
\colhead{Date (UT)} & \colhead{Exp. time [sec]} &
 \colhead{Field}
}
\startdata
2003/03/17  & 18630 & SSDF \\
2003/03/18  & 18000 & SSDF \\
2003/03/19  & 10800 & SSDF \\
2003/03/20  & 18480 & SSDF \\
2003/04/22  & 18630 & SSDF \\
2003/04/23  & 18630 & SSDF \\
            & 135   & M13 (PSF reference)\\
2003/04/24  & 8100  & SSDF \\
            & 135   & M13 (PSF reference)\\
\enddata
\end{deluxetable}

\begin{figure}
\plotone{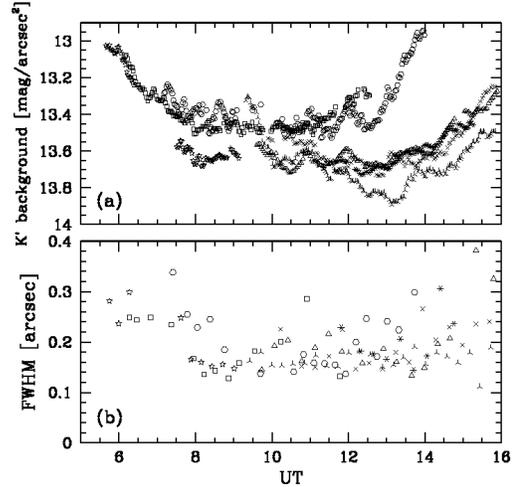}
\caption{The variation of the sky background brightness in the
 $K^{\prime}$ band (a) and the spatial resolution with AO (b) during the
 observations.  Different symbols correspond to different observing
 nights listed in Table \ref{log}. The spatial resolution is represented
 by the FWHM of the point-like source (``S'' in Figure
 \ref{ssdf_image}). \label{obs_condition}}
\end{figure}

Throughout each observing night, the variation of the sky background
brightness was small except at large airmass. The sky variation
corresponds to about $\pm$0.1 mag in the limiting magnitude variation of
each frame (Figure \ref{obs_condition}a), suggesting that the observing
condition was almost stable throughout all observing nights. The
correction performance of AO changes every moment depending largely on
the observational condition. We monitored the variation of spatial
resolution by measuring the FWHM of a point-like source which has the
sharpest and nearly circular profile in our field of view (marked ``S''
in Figure \ref{ssdf_image}) except for the saturated bright stars. Since
this point-like source ``S'' is too faint to be detected in each single
frame, we combined each set of nine-point dithering to measure the
FWHM. The variation of FWHM after AO correction was found to be small
for all frames (Figure \ref{obs_condition}b). Thus, the correction
performance of AO was stable and the point spread function (PSF)
remained to be constant for each night.

\section{Data reduction}
We reduced the data with IRAF\footnote{IRAF is distributed by the
National Optical Astronomy Observatories, which are operated by the
Association of Universities for Research in Astronomy, Inc., under
cooperative agreement with the National Science Foundation.} software
packages. Before the data reduction, we checked all data frames visually
and removed some low-signal frames due to cirrus as well as some frames
with unusual dark patterns due to unstable detector temperature at the
beginning of continuous exposures. We rejected these frames ($\sim$ 15\%
of all frames) before the data reduction and the resultant total
integration time of all used frames is about 26.8 hours. Flat fielding
was performed using the median combined sky flat frame. The sky flat
frames were created for each night using the dark subtracted raw
frames. In generating the sky flat frames, bright and extended objects
in each raw frame were masked in order to reduce their contribution to
the final flat frame.  We made the position map of the bad pixels on the
detector from the dark and flat frames, which were used for picking the
pixels with extraordinarily high counts and the pixels with extremely
low sensitivity, respectively. Bad pixels on our observed frames were
removed by interpolation using this position map. The nine frames of
each dithering set were median-combined to create the sky frame for each
set. Before this process, all of the discernible objects are removed
with interpolation by adjacent sky counts to reduce the contribution of
the individual objects to the resultant sky level. The sky subtraction
was performed for each set of nine-point dithering (every 810 or 1080
seconds) to minimize the effect of time variation of the sky
background. Then, we shifted and combined all of the sky subtracted
frames with exposure time weighting. The image offsets were determined
at a sub-pixel level from the bright objects in each frame.  The final
$K^{\prime}$ image is shown in Figure \ref{ssdf_image}. The average FWHM
of final image was about 0\farcs18 for the relatively bright
($K^{\prime} \sim 21.2$) point-like source ``S''. Throughout this
paper, we used the profile of this point-like source as the PSF of our
AO image.

\section{Source detection and photometry}
\begin{figure*}
\plottwo{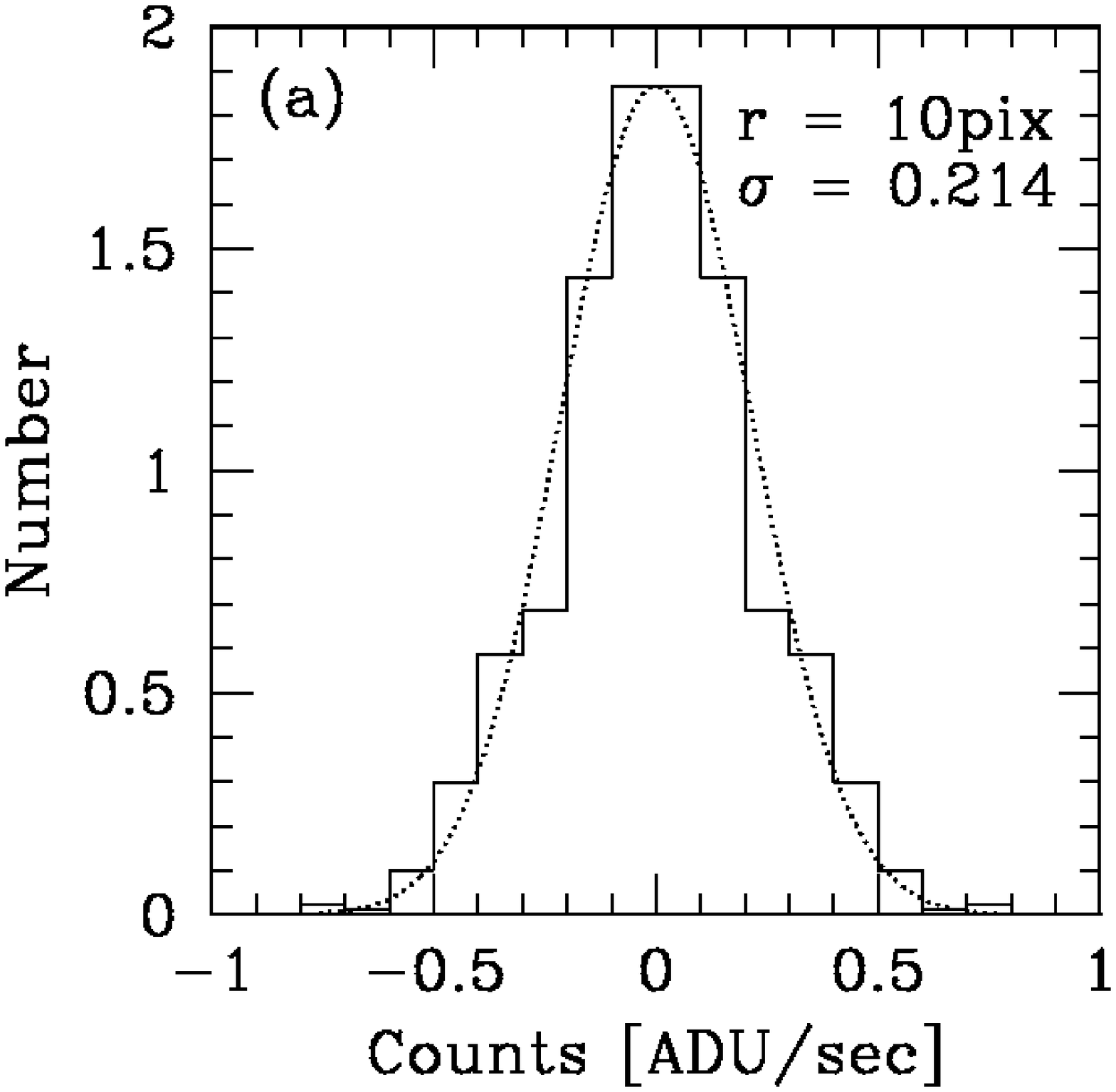}{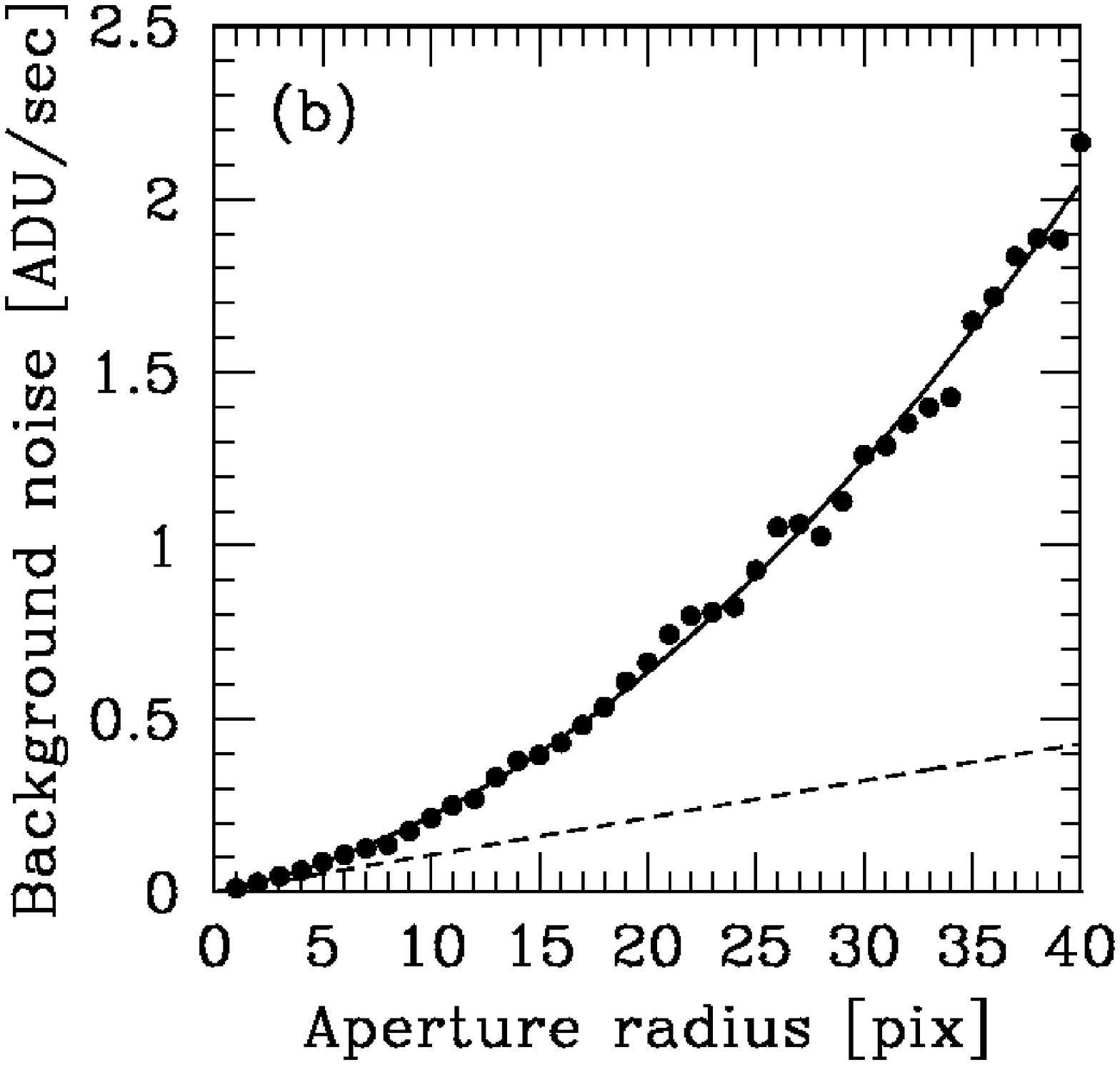} 
\caption{Measurement of background noise within the aperture. (a)
Gaussian function (dotted line) is fitted to the histogram of measured
flux in the aperture area $A = \pi r^2$ ($r=10$ pixels) which was
randomly placed avoiding the object locations. We regarded the width
$\sigma$ of gaussian function as the background noise within the
aperture. This method allows us to measure the true background noise in
which a pixel-to-pixel correlation is taken into account. (b) Relation
of aperture radius $r$ versus background noise within the
aperture. Filled circles show the measured background noise and the
solid line shows the fitted line to the data with a two dimensional
function, $a r^2 + b r$, where $a$ and $b$ are the free parameters. The
measured background noise with large aperture is significantly larger
than that expected from a linear scaling of the pixel-to-pixel noise
(dashed line), probably due to the correlated fluctuations of the
background on large scale. \label{r_noise}}
\end{figure*}%

Source detection and photometry were performed by the SExtractor
\citep{ber96}. Before the source detection, the final image was smoothed
by a gaussian filter ($\sigma$ = 1 pix) to give an optimal source
detection capability with less spurious detection. We defined the
detection threshold as the 1.5$\sigma$ level of the surface brightness
fluctuation of the sky (23.64 mag/arcsec$^2$). If more than 19
contiguous pixels have larger counts than the threshold, we regarded it
as a positively detected source. To reduce the risk of spurious
detection, some areas near the bright stars and the known ghosts of
bright stars are excluded from the detection area (see Figure
\ref{ssdf_image}). As a result, the total number of detected sources is
236 within the net detection area of 0.9 arcmin$^2$, although some
spurious detections due to statistical noise may be included in this
number.

\begin{figure}
 \plotone{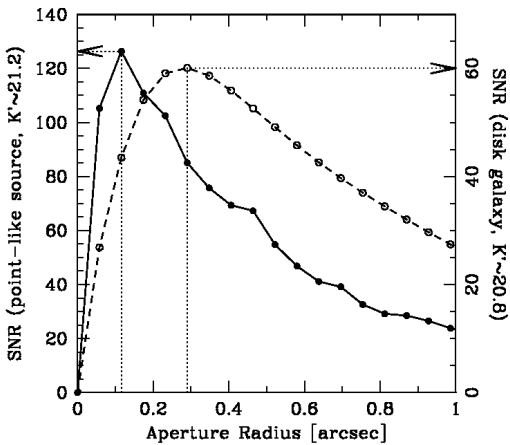}
\caption{Signal-to-noise ratio (SNR) of the $K^{\prime} \simeq 21.2$
point-like source ``S'' (solid line with filled circles, left y-axis)
and the $K^{\prime}\ \simeq\ 20.8$ disk galaxy ``G1'' (dashed line with
open circles, right y-axis) as a function of aperture radius. This
diagram is used to determine the 5$\sigma$ limiting magnitude. Dotted
line shows the maximum SNR and the aperture radius. In case of the
point-like source ``S'', maximum signal-to-noise ratio of 127 is
achieved at aperture radius of 0\farcs1, so that the 5$\sigma$ limiting
magnitude for point sources is $21.2-2.5 \log (5/127) \simeq 24.7$ with
an aperture diameter of 0\farcs2. Similarly, the 5$\sigma$ limiting
magnitude for galaxies was estimated at $K^{\prime} \sim 23.5$ with an
aperture diameter of 0\farcs6. \label{snr}}
\end{figure}%

We performed photometry of detected sources in terms of aperture
magnitude, isophotal magnitude, and total magnitude (see \citet{ber96}
for further description of each magnitude). Throughout this paper, we
mostly use the total magnitude except for the discussion of size
distribution in \S 7.2, where we use the isophotal magnitude. The
photometric calibration was carried out using the infrared faint
standard star GSPC P330-E \citep{per98}. We took the image of this
standard star at airmass $\simeq$ 1 in a photometric night and used it
as the photometric reference for the object frames which were taken in
the same night at the same airmass. Then, a bright galaxy ($K^{\prime}
\sim 16.9$) in the field was used as the photometric reference for the
final image. The resultant zero point in the final combined image is
$K^{\prime} \sim 23.645 \pm 0.011$ mag/ADU.

Understanding the noise properties is crucial because the limiting
magnitude and photometric uncertainty rely on them.  The photometric
uncertainty in typical near-infrared deep images is well described by
the poisson noise of the background signal in each pixel.  However, image
processing, such as shift and combination, has introduced correlations
between neighboring pixels.  If the photometric uncertainty was
estimated from linear scaling of the pixel-to-pixel variation, we would
underestimate it. To estimate the true uncertainty of photometry, we
measured the sky fluxes in several tens of circular apertures at random
position in the final image and regarded the standard deviation $\sigma$
of the measured fluxes as the photometric uncertainty. We changed the
radius of circular apertures ranging from 1 to 40 pixels
(0\farcs058$-$2\farcs3) to cover all the photometric apertures of
detected sources (Figure \ref{r_noise}).  We fitted $\sigma$ with a
two-dimensional function of the aperture radius, $\sigma (r) = ar^2 +
br$, where $a$ and $b$ are the fitting parameters.  We used this
function for calculating the photometric uncertainty of each detected
sources with various apertures.  The measured uncertainty with this
method largely exceeds the uncertainty expected from linear scaling of
the pixel-to-pixel noise (dashed line in Figure \ref{r_noise}). The
large discrepancy at large aperture was also reported in
\citet{lab03}. This could be caused by correlated fluctuations of the
background on large spatial scale.

We estimated the limiting magnitude for point sources using the
point-like source ``S'' in the final image. The limiting magnitude was
determined by using a diagram that shows signal-to-noise ratio (SNR) of
the point-like-source ``S'' as a function of an aperture radius (solid
line in Figure \ref{snr}), that was represented as a ratio of flux to
photometric uncertainty within an aperture. The achieved 5$\sigma$
limiting magnitude for point sources is $K^{\prime} \sim 24.7$
(5$\sigma$) with an aperture diameter of 0\farcs2, where the aperture
size was determined to have the maximum SNR. This is the faintest
$K$-band limiting magnitude for point sources achieved to date. Because
galaxies would have more extended profile than point sources, the
limiting magnitude for galaxies should be brighter than that for point
sources.  Thus, we also estimated a limiting magnitude for galaxies with
the similar method as for point sources using a galaxy ``G1'' (see
Figure \ref{ssdf_image}), that is a disk galaxy which has the typical
effective radius ($r_e = 0\farcs25$) in the SSDF (see \S 6.2 for
details). The achieved 5$\sigma$ limiting magnitude for galaxies is
$K^{\prime} \sim 23.5$ (5$\sigma$) for 0\farcs6 aperture. The number
of detected sources brighter than the limiting magnitude for point
sources is 145, while the total number of detected sources is 236 with
the faintest magnitude of $K^{\prime} \sim 26.0$.

\section{Analysis}

\subsection{Detection completeness}
\begin{figure}
\plotone{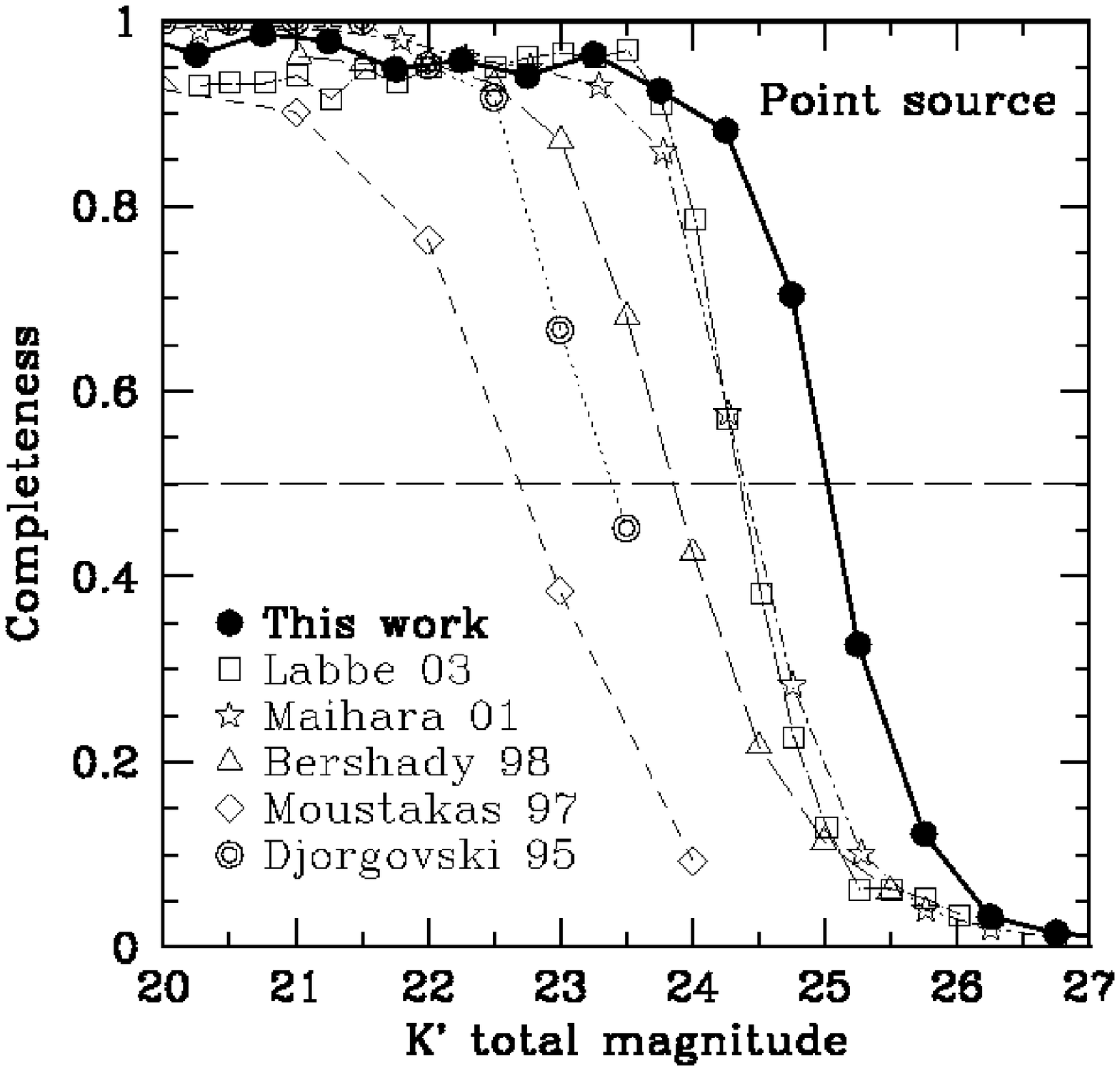}
\caption{Detection completeness curves derived from the simulations
using the artificial point sources. Thick solid line with filled circles
shows the detection completeness for point source of our $K^{\prime}$ survey
in the SSDF, while thin lines with other symbols show that of deep $K$
or $K^{\prime}$ imaging surveys from the literature
\citep{lab03,mai01,ber98,mou97,djr95}. Horizontal dashed line indicate the level of 50\%
completeness. \label{completeness_star}}
\end{figure}

\begin{figure}
\plotone{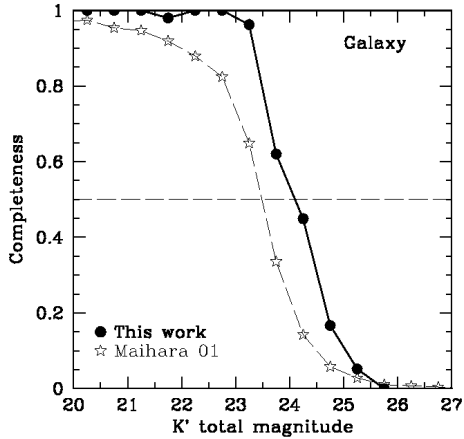}
\caption{Detection completeness curves
derived from the simulations using the artificial galaxies that have the
 typical size in the SSDF. Thick solid line with filled circles shows the
detection completeness for galaxies of our $K^{\prime}$ survey in the SSDF,
while thin dashed line with stellate symbols show that of \citet{mai01}
for a comparison. Horizontal dashed line indicate the level of 50\%
completeness. \label{completeness_gal}}
\end{figure}

At the fainter magnitude, the detection completeness decreases because
of the statistical noise. We estimated the detection completeness for
point sources by conducting the same source detection as described in \S
4 but with a large number of artificial point sources placed at random
positions in the final image. The artificial point sources were created
from the point-like source in the final image (source ``S'' in Figure
\ref{ssdf_image}) by applying a flux scaling in the range of $20 <
K^{\prime} < 27$.  Thick solid line with filled circles in Figure
\ref{completeness_star} shows the detection completeness curve for point
sources in our final image as a function of $K^{\prime}$ magnitude. We
estimated that 50\% completeness for point sources is achieved at
$K^{\prime} \simeq 25.0$, which means a half of all galaxies can be
detected at this magnitude, and it is about 0.6 magnitude deeper than
previous deep imaging such as \citet{lab03} and \citet{mai01}. Thus, our
data should offer the most reliable source detection down to 25 mag.  We
also estimated the detection completeness for galaxies with artificial
galaxies that have the typical size in the SSDF. The artificial galaxies
were created to have the same surface brightness profile as the galaxy
``G1'', which was used in the limiting magnitude estimation in \S4. We
placed them at random positions with varied flux in the range of $20 <
K^{\prime} < 27$. Figure \ref{completeness_gal} shows the estimated
detection completeness curve for galaxies. The completeness curve for
galaxies in our final image are shown as thick solid line with filled
circles. For a comparison, the completeness curve for extended source in
the previous deep imaging by \citet{mai01} is also shown as thin dashed
line with stellate symbols in the same diagram. We estimated that 50\%
completeness for extended source is achieved at $K^{\prime} \simeq
24.1$, that is about 0.6 mag fainter than that of \citet{mai01}. This
magnitude gain is same as for the gain for point source, suggesting that
the AO system improves the efficiency of detecting high-redshift
galaxies as well as that of point sources.  In this paper, the detection
completeness for point source is used to correct the number
counts. Therefore, the detection completeness here should be considered
upper limits and then the corrected counts should be considered lower
limits.

\subsection{Noise contamination to number counts}
\begin{figure}
\plotone{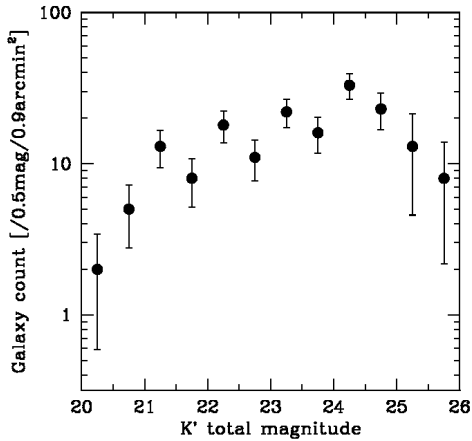}
\caption{Galaxy counts in the SSDF in the 0.5 magnitude bin without any
 correction except for subtraction of the noise contamination. Error bars
 show the uncertainties due to the poisson statistics.
 \label{numcnt_ssdfraw}}
\end{figure}

We detected the objects down to $K^{\prime} < 26$ in our final image.
Based on the photometric measurements of these objects, we derived the
number counts in the 0.5 magnitude bin (Table \ref{count}). The number
counts, in principle, contain not only the counts from galaxies, but
also the counts from the Galactic stars and spurious detections due to
the effect of statistical noise. The contribution of the Galactic stars
to our number counts should be negligible, because the expected star
counts toward the Galactic pole are less than 0.2
mag$^{-1}$acrmin$^{-2}$ in all magnitude range \citep{min98b,fug03}. On
the other hand, the contribution of spurious detection is significant in
the faint end, where the signal-to-noise ratio for the detection becomes
lower. We estimated the number of spurious detections from an artificial
noise frame which was created by combining the sky subtracted frames
without adjusting the dithering offset. The combination was conducted
with the options of exposure time weighting, which is similar to the
combination of the final image. Before the combination, all discernible
objects in each image were removed by interpolation and the estimated
noise from the adjacent pixels was added on the interpolated
pixels. Even after removing the discernible objects, even fainter
objects could be remained in the noise in each single frame. In order to
minimize the contribution of the light from such faint objects, the sign
of each image was reversed before the combination. We applied the same
detection procedure to the artificial noise image as used for the source
detection. The resultant number of spurious detections in each 0.5
magnitude bin is given in Table \ref{count}. We derived the galaxy
number count in each magnitude bin by subtracting the number of spurious
detection from the raw count. The estimated galaxy counts are listed in
Table \ref{count} and shown in Figure \ref{numcnt_ssdfraw}.

\begin{deluxetable*}{ccccccccc}
\tabletypesize{\scriptsize}
\tablecaption{The corrected  $K^{\prime}$-band galaxy number counts. \label{count}}
\tablewidth{0pt}
\tablehead{
\colhead{$K^{\prime}$} & \colhead{$n_{raw}$}  & \colhead{$n_{noise}$} &
 \colhead{$n_{gal}$}& \colhead{$\Delta n_{gal}^{p}$} & \colhead{$\Delta
 n_{gal}^{p+c}$} & \colhead{Completeness} & \colhead{$N_{cor}$} &
 \colhead{$\Delta N_{cor}$}\\
\colhead{(1)} & \colhead{(2)} & \colhead{(3)} & \colhead{(4)} &
 \colhead{(5)} & \colhead{(6)} & \colhead{(7)} & \colhead{(8)} &
 \colhead{(9)}
}
\startdata
20.25 & 2  & 0  & 2  & 1.41 & 1.51  & 0.96 & 1.959E4 & 1.287E4 \\
20.75 & 5  & 0  & 5  & 2.24 & 2.59  & 0.99 & 4.837E4 & 2.157E4 \\
21.25 & 13 & 0  & 13 & 3.61 & 4.97  & 0.98 & 9.735E4 & 4.168E4 \\
21.75 & 8  & 0  & 8  & 2.83 & 3.54  & 0.95 & 9.112E4 & 3.067E4 \\
22.25 & 18 & 0  & 18 & 4.24 & 6.39  & 0.96 & 1.360E5 & 5.479E4 \\
22.75 & 11 & 0  & 11 & 3.32 & 4.44  & 0.94 & 1.253E5 & 3.871E4 \\
23.25 & 22 & 0  & 22 & 4.69 & 7.47  & 0.96 & 1.598E5 & 6.367E4 \\
23.75 & 17 & 1  & 16 & 4.24 & 6.06  & 0.92 & 2.100E5 & 5.377E4 \\
24.25 & 37 & 4  & 33 & 6.40 & 11.15 & 0.88 & 2.497E5 & 1.038E5 \\
24.75 & 31 & 8  & 23 & 6.24 & 9.47  & 0.70 & 2.577E5 & 1.104E5 \\
25.25 & 42 & 29 & 13 & 8.43 & 10.30 & 0.33 & 3.038E5 & 2.587E5 \\
25.75 & 21 & 13 & 8  & 5.83 & 8.32  & 0.12 & 4.746E5 & 5.575E5 \\
\enddata
\tablecomments{(1) $K^{\prime}$-band total magnitude. (2) Raw counts of
 detected objects in the 0.5 magnitude bin. (3) Noise counts in the 0.5
 magnitude bin. (4) Galaxy counts in the 0.5 magnitude bin ($n_{gal} =
 n_{raw} - n_{noise}$). (5) Uncertainties in the galaxy counts coming
 from poisson statistics. (6) Uncertainties in the galaxy counts coming from
 poisson statistics and clustering of galaxies. (7) Detection
 completeness for point source. (8) Differential galaxy number counts in
 mag$^{-1}$deg$^{-2}$ corrected for the incompleteness and the scatter
 due to the photometric uncertainties. (9) Uncertainty in the corrected
 differential number counts coming from the poisson statistics and the
 clustering of galaxies.}
\end{deluxetable*}

\subsection{The scatter of number counts due to photometric uncertainty}
Even if the noise counts are removed, the scatter of galaxy number
counts among magnitude bins still remains due to photometric
uncertainty. In order to evaluate the effect of this scatter, we
performed a simulation by adding artificial objects to the final image
and applying the same detection procedure as described in \S 4. In this
simulation, about 10 artificial objects were added 450 times (total of
about 4,500 objects) in order not to largely change the number density
of the objects in the final image and the detection procedure was
performed each time. The artificial objects of various magnitudes were
created from the point source in the final image by scaling its
flux. The magnitudes of the artificial objects were assumed to
distribute with power-law, $N(m) \propto 10^{\alpha m}$, where $N$ is
the number of the artificial objects in each 0.5 magnitude bin, $m$ is
the central magnitude of each magnitude bin, and $\alpha$ is the
power-law index. We performed the simulation only with point sources
because previous analyses related to the number counts are based on
point sources to give the lower limit of the counts. An initial
power-law index of ${\alpha}_0 = d\log N/dm$ was assumed to be the slope
of raw galaxy counts over a reliable magnitude range, $22 < K^{\prime} <
24$, where the signal-to-noise ratio is greater than 3 (i.e. more than
10 raw counts) and the detection completeness for point sources is
higher than 90\%.

\begin{figure}
\plotone{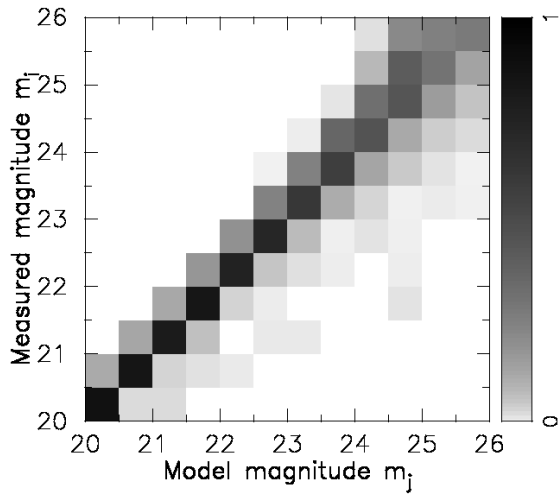}
\caption{The probability matrix for detected galaxies in the SSDF based
 on the simulation using the artificial point sources. Each
 element of the matrix gives the probability that a galaxy detected with
 $m_i$ actually has a model magnitude of $m_j$. The probability is color-coded
 according to the scale bar on the right of this figure. \label{pmatrix}}
\end{figure}

Based on the simulation, we first generated the transfer matrix
$T_{ij}$, each element of which gives the fraction of galaxies with
magnitude $m_j$ but detected with $m_i$. Then, we generated the
probability matrix $P_{ji}$, each element of which gives the probability
that an artificial object detected with $m_i$ actually has a magnitude of
$m_j$. These elements are described as
\begin{equation}
 P_{ji} = T_{ij} n_j / \sum_k T_{ik} n_k\ , 
\end{equation}
where $n_j$ is the number of the artificial objects with magnitude
$m_j$. The probability matrix is shown in Figure \ref{pmatrix}. 
With this probability matrix, we corrected the galaxy number counts as  
\begin{equation}
 n^{cor}_j = \sum_i P_{ji} n^{gal}_i\ ,
\end{equation}
where $n^{cor}_j$ is the corrected galaxy number count at magnitude
$m_j$, and $n^{gal}_i$ is the raw galaxy count with detected magnitude
$m_i$ (Table \ref{count}). If the slope of corrected galaxy counts was
different from the initial slope ${\alpha}_0$ used in the simulation,
the correction procedure described above was repeated with different
initial slope until the corrected slope coincides with the initial
slope. This method has been employed in some studies of galaxy number
counts \citep{sma95,min98a,mai01}.

\begin{figure}
\plotone{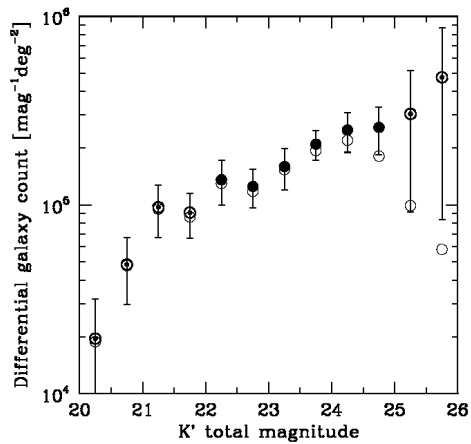}
\caption{Differential $K^{\prime}$-band number counts of galaxies
estimated from the galaxy counts in the SSDF.  Filled circles show the
galaxy counts in a reliable magnitude range of $22 < K^{\prime} < 25$,
where the signal-to-noise ratio for raw counts is greater than 3 and the
detection completeness for point source is higher than 50\%. Circled
dots show the less reliable galaxy counts. These galaxy counts are
corrected for the detection completeness and the scatter due to the
photometric uncertainty. Open circles show the galaxy counts without
completeness correction. The completeness correction becomes significant
at $K^{\prime} > 25$. Error bars show the uncertainties due to the
poisson statistics. \label{numcnt_ssdf}}
\end{figure}

We derived the differential galaxy number counts ($N^{cor}_i$) in the
unit of number mag$^{-1}$ deg$^{-2}$ from the counts $n^{cor}$ after
correcting for the detection completeness estimated for point source in
\S 5.1 (Table \ref{count}). Figure \ref{numcnt_ssdf} shows the plot of
galaxy number counts with and without the completeness correction. At
brighter magnitude ($K^{\prime} < 22.0$), our differential number
counts are not complete with less than 3$\sigma$ measurements (i.e. less
than 9 raw counts) because of a small field of view (1\arcmin$\times$
1\arcmin). At fainter magnitude, the detection completeness for point
source becomes lower than 50\% at $K^{\prime} > 25$. Thus, the
reliable magnitude range of our differential number counts is $22 <
K^{\prime} < 25$.

\subsection{Field-to-field variations of galaxy number counts}
Clustering of galaxies could lead to the systematic uncertainties in the
number counts. We estimated the field-to-field variations of the counts
due to the clustering from an angular correlation function. Considering an
angular area of $\Omega$, with a mean count of $\langle N \rangle$
galaxies, the variance in the number of detected galaxies is
\begin{equation} 
 \mu_2 = \langle N \rangle + \frac{\langle N \rangle^2}{\Omega^2} \int \int \omega (\theta_{12})
  \,d\Omega_1 d\Omega_2\ ,
  \label{clustering}
\end{equation}
where $\omega(\theta)$ is the angular correlation function of galaxies
and $\theta_{12}$ is the angle between the solid-angle elements
$d\Omega_1$ and $d\Omega_2$ \citep{gro77}. In this formula, first term
comes from poisson statistics and second term comes from the clustering.
The angular correlation function for faint galaxies has been measured in
the $K$ band in a number of literatures (e.g.,
\citealt{bau96,car97,roc99b}), and it is described in a form of $\omega
(\theta) = A\theta^{-0.8}$. The amplitude $A$ at fixed $\theta$ is a
decreasing function of $K$ magnitude, while there is an evidence that
the amplitude becomes relatively flat at $K > 20$ with a value of $A
\sim 1.1 \times 10^{-3}$ when $\theta$ is measured in degree
\citep{roc99b}. In this paper, we assume the amplitude-magnitude
relation does not turn over at $K > 20$, which is theoretically
reasonable (e.g., \citealt{roc99a}), to set the upper limit on the
uncertainty of the counts coming from the clustering of galaxies. For the
0.9 arcmin$^2$ area of the SSDF, the variance becomes $\langle N \rangle
+ 61.3 A \langle N \rangle^2$. The estimated uncertainties of the galaxy
counts coming from poisson statistics and clustering at each magnitude
bin are summarized in Table \ref{count}. The uncertainties due to
poisson statistics and clustering are not much larger than that only due
to poisson statistics.

\begin{figure}
\plotone{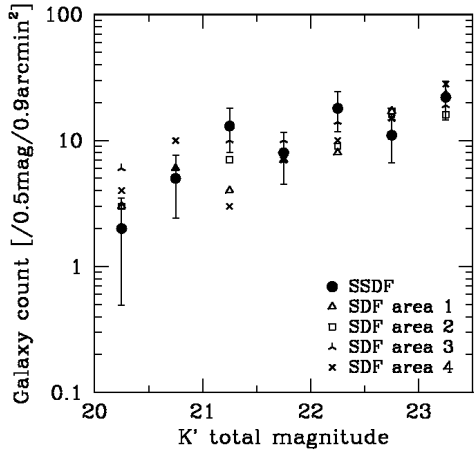}
\caption{The galaxy counts in the SSDF with the uncertainties due to the
 poisson statistics and the clustering of galaxies estimated from the
 angular correlation function for the 0.9 arcmin$^2$ area of the SSDF
 (filled circles). The galaxy counts for four discrete 0.9 arcmin$^2$
 areas in the SDF are also plotted (open circles). The estimated
 uncertainties of the SSDF due to a
poisson statistics and a clustering are almost consistent with the
variance of the galaxy counts for four areas in the SDF.  \label{field_var}}
\end{figure}

In order to check the validity of the variation estimated from the
angular correlation function, we also estimated the field-to-field
variations of the counts using the SDF data \citep{mai01}, which has
almost four times larger area (1\farcm97 $\times$ 1\farcm9) than that of
the SSDF.  We derived the counts from four discrete 0.9 arcmin$^2$ areas
in the SDF at $20 < K^{\prime} < 23$ where the detection completeness is
higher than 90\% for both of the SDF and the SSDF.  The galaxy counts of
the SSDF are plotted in Figure \ref{field_var} with the uncertainties
estimated from equation (\ref{clustering}). The galaxy counts for four
discrete areas in the SDF are also plotted.  It is found that the
uncertainties due to the poisson statistics and the clustering of
galaxies are almost consistent with the variations of the galaxy counts
from four areas in the SDF.

\section{Performance of AO deep imaging}
\subsection{Comparison with seeing-limited observation}
To estimate the performance of our AO deep imaging, we compared
the quality of our AO image of the SSDF to that of the seeing-limited
image of the SDF \citep{mai01}. The SDF data were obtained by the Subaru/CISCO
under the FWHM$\sim$0\farcs45 seeing condition with the total
integration time of 9.7 hours. Because a part of the SDF field of view
is included in the SSDF (see Figure \ref{field}), we can make a
direct comparison between AO and seeing-limited data.

\begin{figure}
\plotone{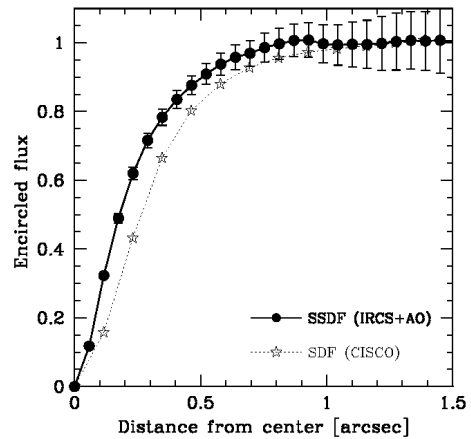}
\caption{Encircled flux of the point-like source ``S'' in the SSDF
(filled circles) as a function of distance from the center of the
stellar image. For a comparison, encircled flux of a point source in the
SDF \citep[stars,][]{mai01} is also shown. Since there is no point
source in the overlapped region between the SSDF and the SDF, a
different point source is used for the SDF. Total flux was normalized to
unity for both point sources. Vertical error bars show the photometric
uncertainties in each aperture.  Because the point-like source ``S'' in
the SSDF is much fainter than the point source used for the SDF, the
photometric uncertainties for the SSDF data are much larger than that
for the SDF data. Much higher concentration of flux was attained in the
SSDF compared to the SDF. \label{encircle}}
\end{figure}

Figure \ref{encircle} shows the encircled flux plot of the point sources
detected in the SSDF and the SDF. The 50\% of the total flux is
contained within a radius of 0\farcs18 for the SSDF, while 0\farcs27 for
the SDF. The higher flux concentration due to the AO correction resulted
in much higher sensitivity. To estimate the sensitivity gain for our AO
observation against seeing-limited observation, we derived the 5$\sigma$
limiting magnitude for point sources in the SSDF and in the SDF with the
same method as described in \S 4. The 5$\sigma$ limiting magnitude is
about 22.9 (0\farcs2 aperture) for the SSDF and 22.4 (0\farcs4 aperture)
for the SDF with an hour integration time. Thus, the sensitivity gain
due to AO correction is about 0.5 mag in the $K^{\prime}$ band.

\begin{figure*}
 \plottwo{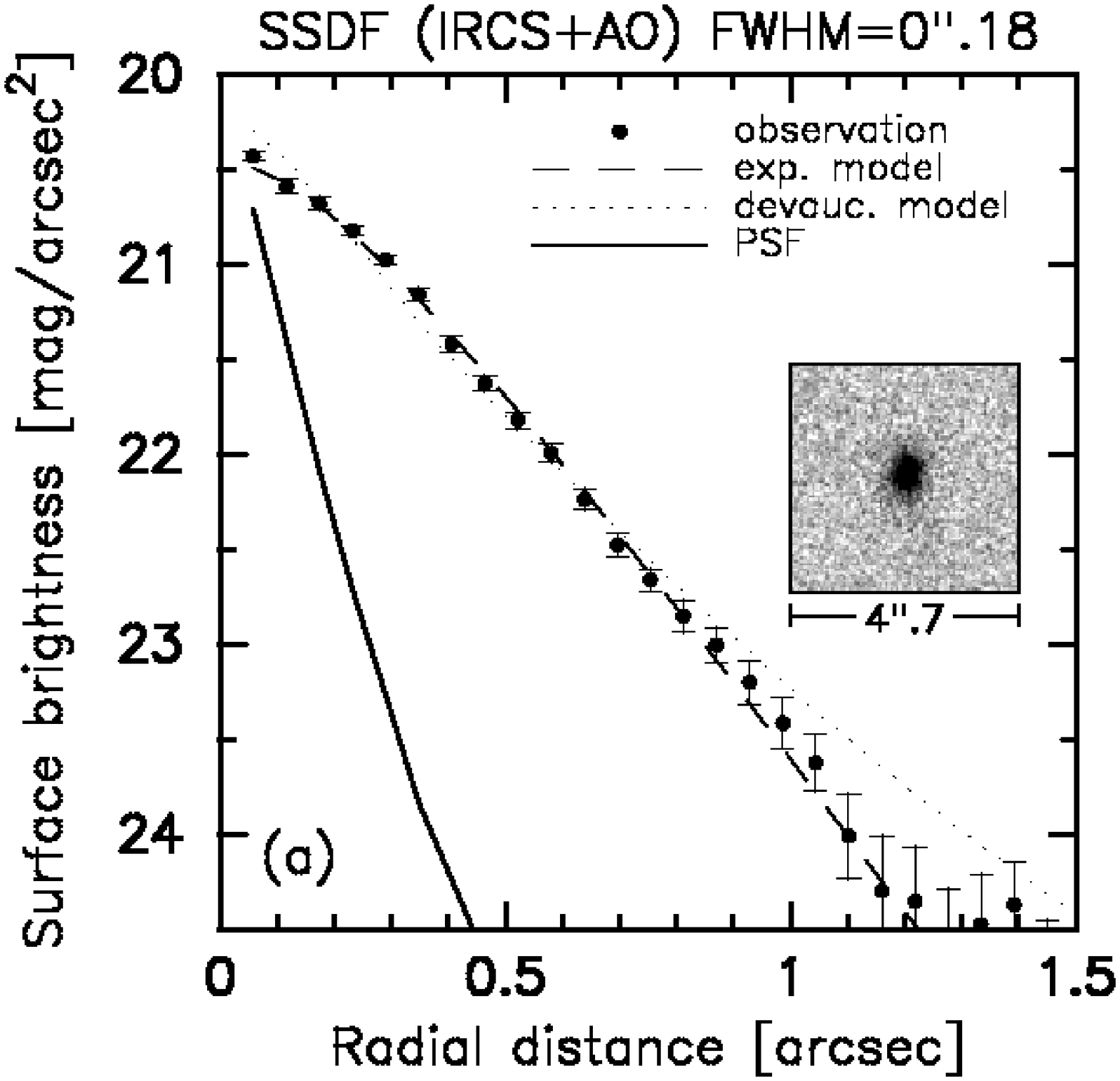}{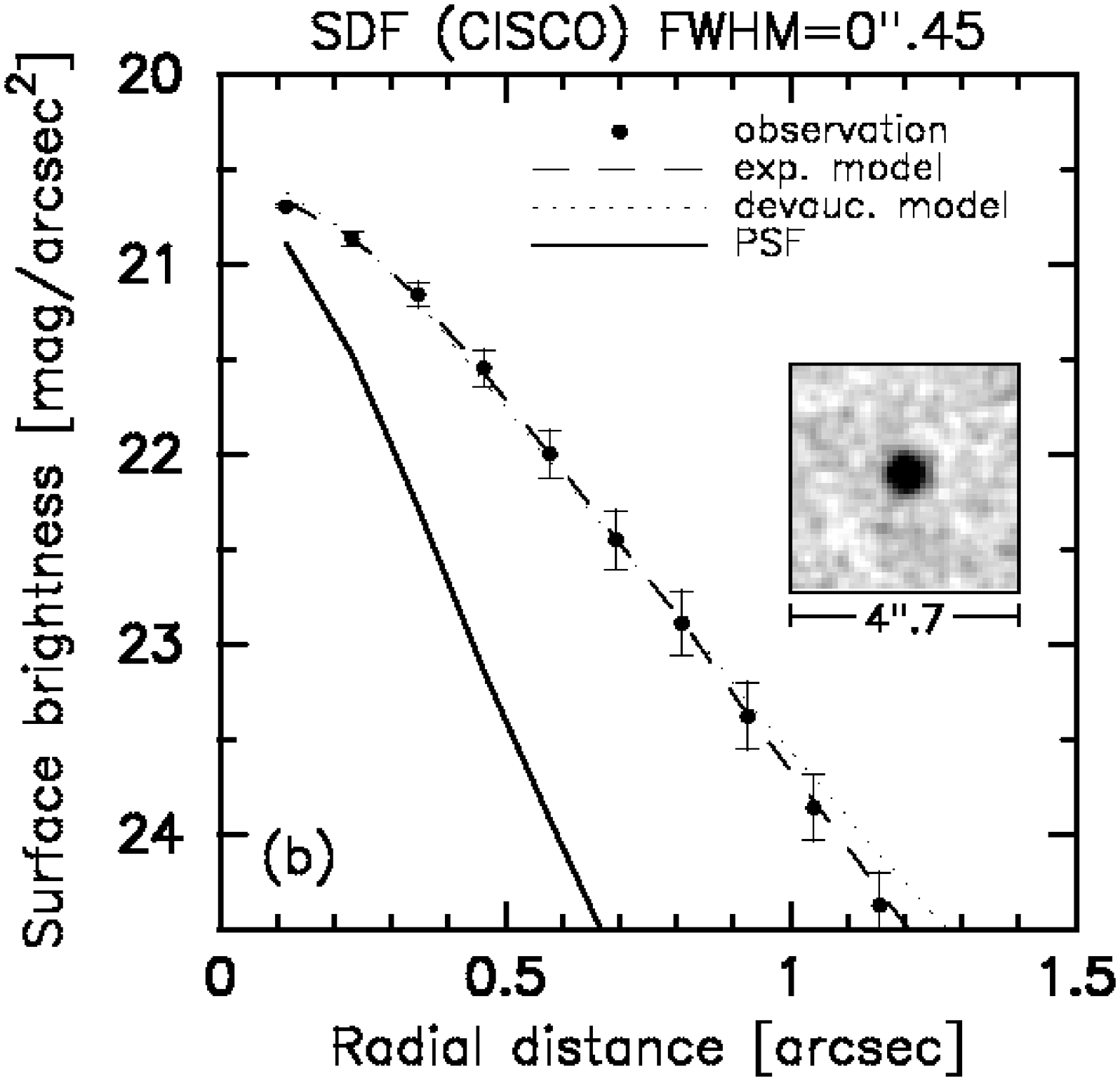}
\caption{Radial profile fitting to a $K^{\prime} \sim 21.3$ galaxy at $z
\sim 1.46$ (``G2'' in Figure \ref{ssdf_image}) in (a) the SSDF with
IRCS+AO and (b) the SDF with CISCO \citep{mai01}. Filled circles show
the observed data in both figures. The PSF-convolved de Vaucouleurs
(dotted line) and exponential (dashed line) profiles were used to model
a galaxy dominated by bulge and disk, respectively. Solid lines show the
PSF profile in both figures. The high-resolution (FWHM = 0\farcs18) image of
the SSDF clearly shows that the best fitted model is the exponential
(disk dominated) profile, while it is difficult to distinguish the two
models with the seeing-limited image of the SDF because of the large
 FWHM  (= 0\farcs45). \label{radfit}}
\end{figure*}%

The improvement of spatial resolution with AO is also significant. Figure
\ref{radfit} shows a model fit to the surface brightness profile of a
$K^{\prime} \sim 21.3$ galaxy at $z \sim 1.46$ (``G2'' in Figure
\ref{ssdf_image}), using a PSF-convolved de Vaucouleurs profile
\citep{dev48} for a bulge-dominated galaxy and an exponential profile
\citep{fre70} for a disk-dominated galaxy.  While it was difficult to
distinguish these model profiles for the seeing-limited SDF data, this
galaxy is evidently a disk dominated galaxy for our SSDF
data. Effective radii of local galaxies, except for compact dwarf
galaxies, range from $\sim$1.0 kpc to $\sim$ 10 kpc depending on their
luminosity \citep{ben92,imp96}. Since our spatial resolution of
0\farcs18 corresponds to about 1.4 kpc at $z\sim1$, we can determine the
size as well as the morphological type even at $z>1$. Thus, our AO
morphology data should enable systematic and quantitative study of
rest-frame optical morphology of galaxies at $z>1$ (Minowa et al. 2005,
in preparation).

\subsection{Effect of partially corrected PSF}
\begin{figure}
\plotone{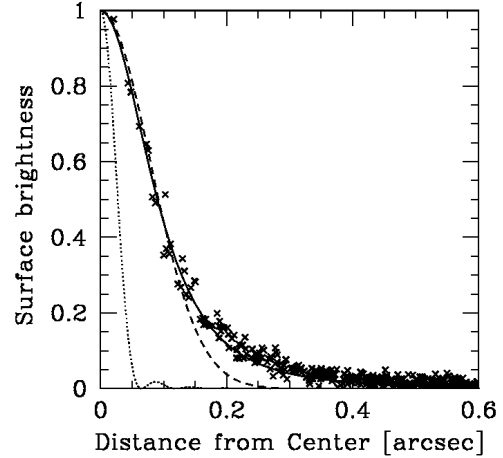}
\caption{Radial profile of the observed PSF in the SSDF. The peak
 surface brightness of the PSF is normalized to unity. The moffat
 profile (solid line) and the gaussian profile (dashed line) are fitted
 to the observed PSF. The moffat profile represents well the observed
 PSF, while the gaussian profile only represents the central part of the
 observed PSF and failed to represents the extended wings. The observed
 PSF has a broader FWHM than the Subaru diffraction-limited PSF (dotted
 line) because of the low Strehl ratio ($\sim$ 0.1).  \label{psf_prof}}
\end{figure}

Typical AO system operated with a guide star of moderate brightness can
only partially correct for turbulence-induced distortions. The partially
corrected PSF consists of two components: a diffraction-limited core
superimposed on a seeing halo. As the Strehl ratio becomes lower, which
means a decrease of the degree of correction, a greater proportion of
energy in the diffraction-limited core is scattered into the surrounding
halo and the peak of the diffraction-limited core becomes lower. As a
result, the seeing halo becomes the dominant component in the PSF. For
the present SSDF data with the Strehl ratio of about 0.1, since about
90\% of the energy is scattered into the surrounding halo, the FWHM of
the observed PSF ($\sim 0\farcs18$) is broader than the
diffraction-limited core ($\sim 0\farcs057$ in the $K$-band) and the
wings extended out to the seeing size are seen around the edge of the
PSF (see Figure \ref{psf_prof}).

\begin{figure}
\plotone{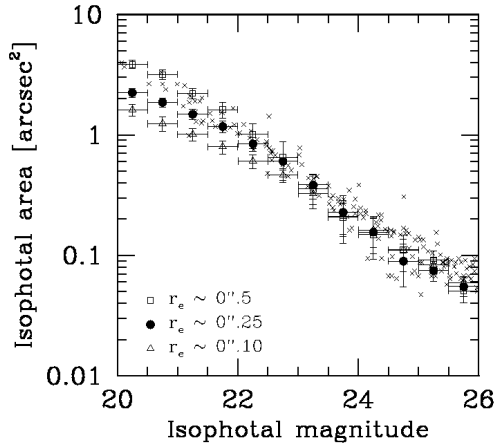}
\caption{The isophotal area versus isophotal magnitude diagram for the
observed galaxies (crosses). We also plotted the mean isophotal area for
the simulated disk galaxies with effective radius ($r_e$) of 0\farcs.1
(open triangles), 0\farcs25 (filled circles), and 0\farcs5 (open
squares). The horizontal error bars show the bin size around the mean
and the vertical error bars show the $1\sigma$ dispersion of the
isophotal area of the simulated galaxies.  The simulated disk galaxies
are convolved with the observed PSF modeled by the moffat profile. The
isophotal areas of the observed galaxies are roughly distributed between
that of the simulated disk galaxies with $r_e = 0\farcs1$ and
$0\farcs5$. Thus, we used the disk galaxy with the intermediate size of
$r_e = 0\farcs25$ as the artificial galaxy that models the typical
galaxies in the SSDF image. \label{mag_iso_obs}}
\end{figure}

The surface brightness profile of galaxies detected in the SSDF may be
distorted by the extended wings of the observed PSF.  To estimate the
effect of the extended wings of the observed PSF on the flux and size
measurement of the galaxies in the SSDF image, we conducted a simulation
using artificial galaxies that were convolved with a model PSF
profile. Figure \ref{psf_prof} shows the radial profile of the observed
PSF and the fitted lines with a moffat profile and a gaussian profile.
These profiles are described as
\begin{mathletters}
\begin{eqnarray}
 I(r) & = & \left[ 1 + (2^{1/\beta} -1)
	 \left(2r/FWHM\right)^2\right]^{-\beta}\ \ (moffat), \\
 I(r) & = & \exp\left[ -\ln(2) \left(2r/FWHM\right)^2\right]\ \ (gaussian),
\end{eqnarray}
\end{mathletters}
where $r$ is the distance from the center of the PSF, $FWHM$ is the full
width at half maximum of the PSF and $\beta$ is the constant related to
the shape of the profile. As $\beta$ decreases, the moffat profile
deviates from the gaussian profile and comes to show more extended
wings. For the observed PSF in the SSDF, best fit parameters of $FWHM$
and $\beta$ are 0\farcs18 and 1.73, respectively. We found that the
moffat profile represents well the observed PSF, while the gaussian
profile fails to represents the extended wings of the observed PSF (see
Figure \ref{psf_prof}). Thus, we used the fitted moffat profile to
convolve the artificial galaxies in the simulation. The artificial
galaxies are created to have the typical size in the SSDF image. Figure
\ref{mag_iso_obs} shows an isophotal area of the observed galaxies, that
means the area where the surface brightness of the galaxies is brighter
than the detection threshold of 23.64 mag/arcsec$^2$, as a function of
the isophotal magnitude. We also plotted the isophotal areas of the
simulated disk galaxies with $r_e =$ 0\farcs1, 0\farcs25, and
0\farcs5. Because the sizes of the observed galaxies are roughly
distributed between the size distribution of the simulated disk galaxies
with 0\farcs1 and 0\farcs5, we used the disk galaxy with the
intermediate size of $r_e = 0\farcs25$ as the artificial galaxy that
models the typical galaxies in the SSDF image. The size and
morphological type of the artificial galaxies are same as the galaxy
``G1'', which was used to estimate the limiting magnitude and the
detection completeness for galaxies (see \S4 and \S5.1). After we placed
the artificial galaxies at random positions in the SSDF image with
varying their fluxes, we measured their magnitude and isophotal area
with the same technique as used for the observed galaxies.

\begin{figure}
\plotone{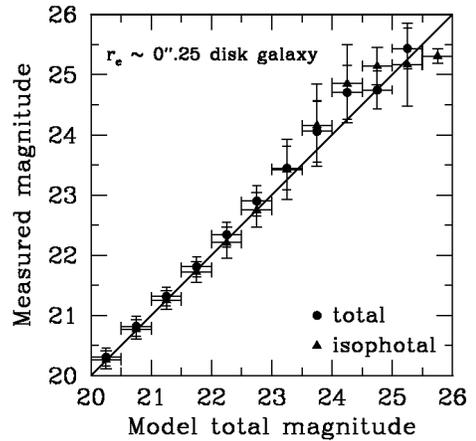}
\caption{Comparison between the model magnitude and the measured
 magnitude of the artificial galaxies that have the typical size in the SSDF
 image (see Figure \ref{mag_iso_obs}). Filled circles show the
total magnitude and filled triangles show the isophotal magnitude. Solid
line shows where the model magnitude is equal to the measured
magnitude. The measured total and isophotal magnitudes are almost
equal to the model magnitude. \label{mag_mag}}
\end{figure}

Figure \ref{mag_mag} shows a comparison between the model magnitude and
the measured magnitude (total and isophotal) of the artificial galaxies.
The measured total and isophotal magnitudes are almost consistent with
the model magnitude within the uncertainties, although the measured
magnitude of the artificial galaxies at the faint end tend to be
slightly fainter than the model magnitude. Thus, the extended wings of
the observed PSF should not affect the measurement of the magnitude of
the typical galaxies in the SSDF image.

\begin{figure*}
\plottwo{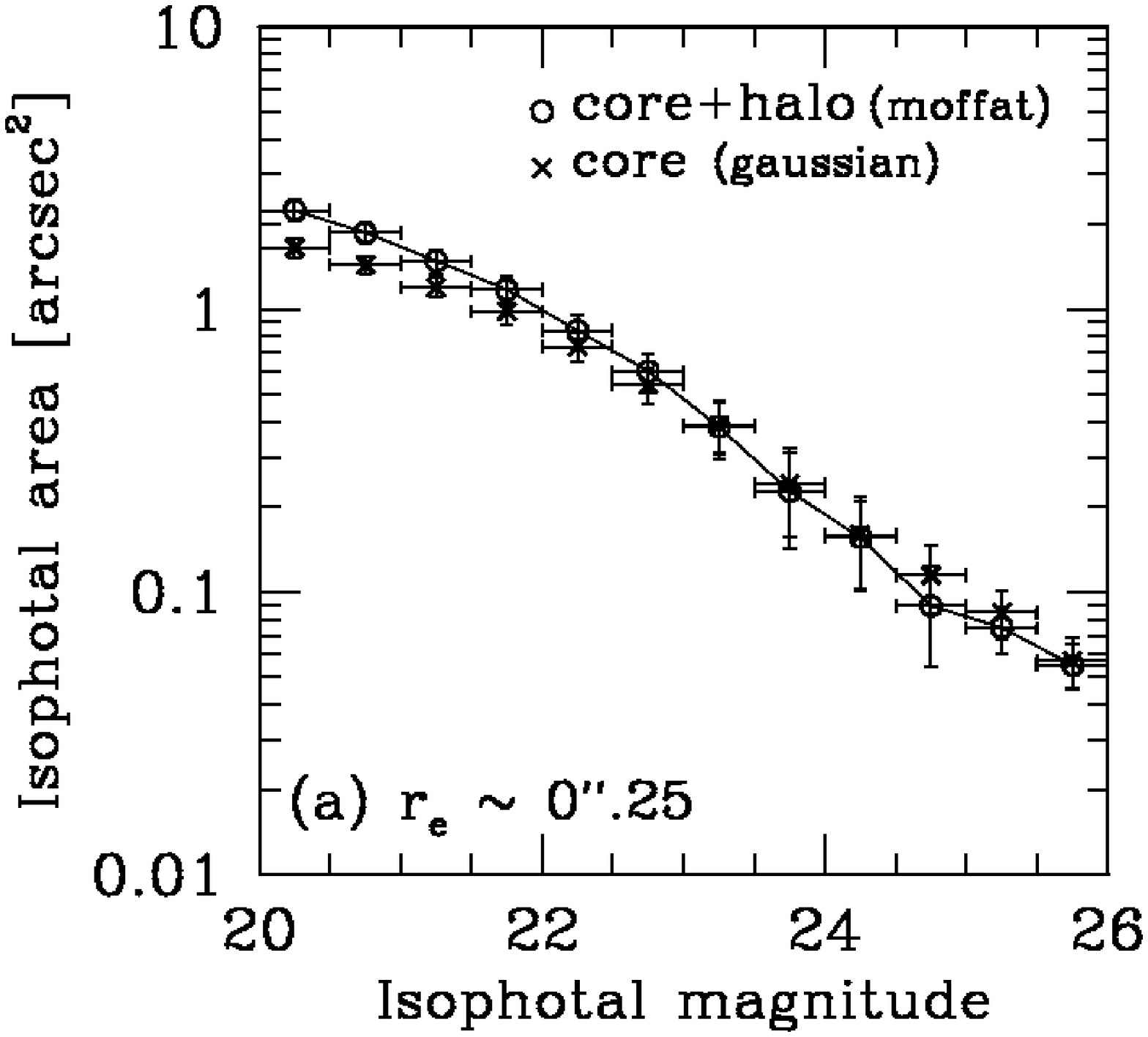}{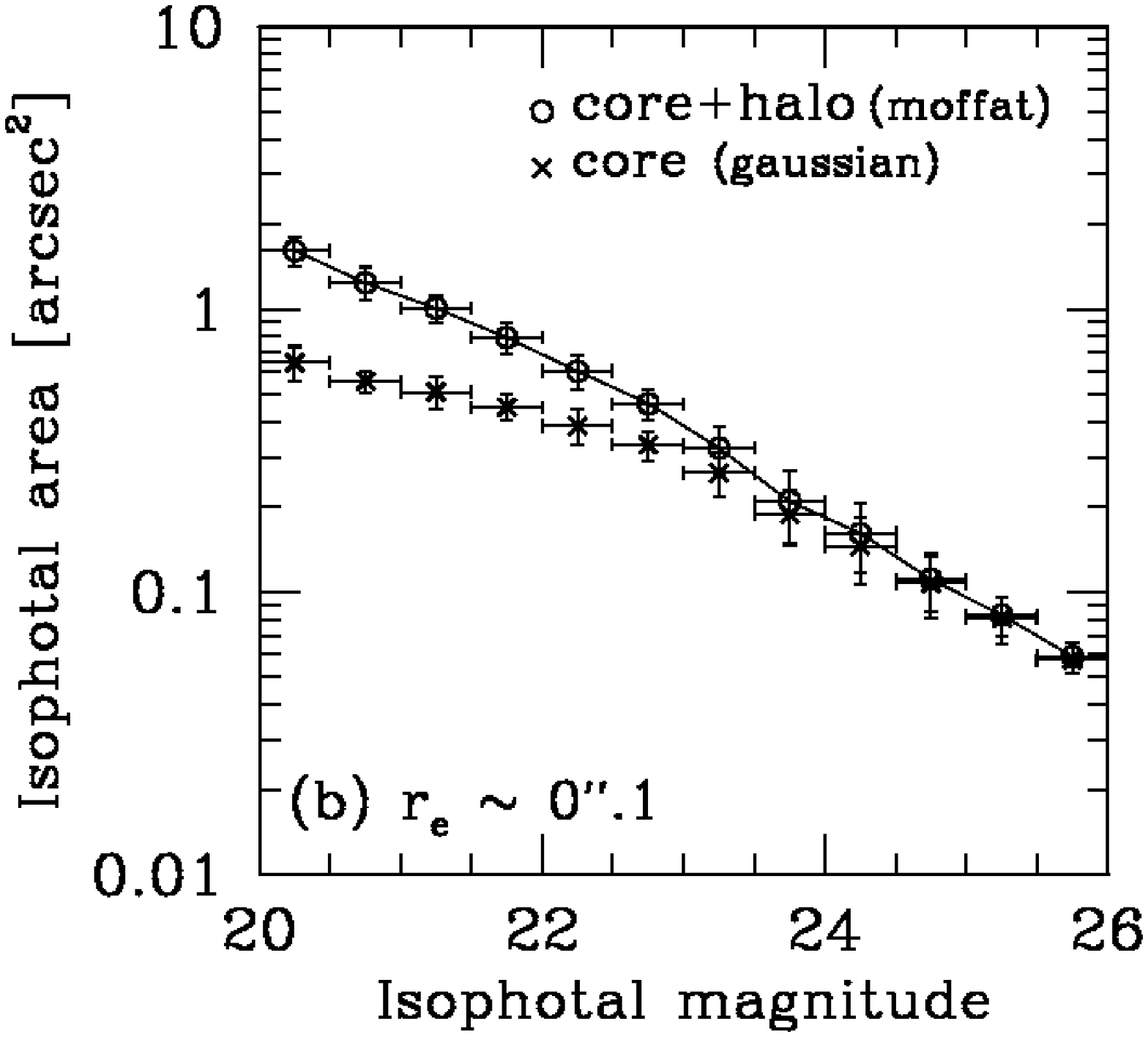}
\caption{The isophotal area of the artificial galaxies are shown as a
function of their isophotal magnitude ($m_{iso}$). Left panel (a) shows
the case for disk galaxies with effective radius ($r_e$) of 0\farcs25,
which is the typical size in the SSDF. We also show the same diagram for
smaller disk galaxies with $r_e = 0\farcs1$ in right panel (b) for a
comparison. The artificial galaxies are convolved with a moffat profile
(open circles) and a gaussian profile (crosses). The moffat profile
represents well the observed PSF that have extended wings, while the
gaussian profile only represents the central part of the observed
PSF. The isophotal area of the $r_e = 0\farcs25$ galaxies convolved with
the moffat profile is not much different from that convolved with the
gaussian profile. On the other hand, the isophotal area of the $r_e =
0\farcs1$ galaxies convolved with moffat profile is larger than that
convolved with the gaussian profile at $m_{iso} \leq
23$. \label{mag_iso_simul}}
\end{figure*}

Figure \ref{mag_iso_simul}(a) shows the isophotal area of the artificial
galaxies with $r_e = 0\farcs25$ as a function of the isophotal
magnitude. We also show the isophotal area versus isophotal magnitude
diagram for small galaxies with $r_e = 0\farcs1$ in Figure
\ref{mag_iso_simul}(b) for a comparison. To estimate the effect of the
extended wings of the observed PSF on the galaxy size measurement, we
compared the isophotal area of the artificial galaxies convolved with
the moffat profile and that convolved with the gaussian profile. Both of
the profiles were fitted to the observed PSF. The moffat profile
represents well the observed PSF that have extended wings, while the
gaussian profile only represents the central part of the observed PSF
(see Figure \ref{psf_prof}). The isophotal areas of the $r_e =
0\farcs25$ galaxies convolved with the moffat profile are almost
consistent with those convolved with the gaussian profile at all
magnitude range. On the other hand, the isophotal areas of the $r_e =
0\farcs1$ galaxies convolved with the moffat profile are larger than
that convolved with the gaussian profile. These implies that the
extended wings of the observed PSF should not affect the isophotal area
of the typical galaxies in the SSDF image, while it may affect the
isophotal area of small galaxies at the bright magnitude range.

\subsection{Effect of angular anisoplanatism}
The performance of AO correction degrades gradually with increasing
distance from a guide star. This is due to increased decorrelation of
the turbulence-induced aberration away from the location of the guide
star (angular anisoplanatism, e.g. \citealt{har98}). Therefore, spatial
resolution (thus sensitivity) is usually highest in the region close to
the guide star and decreases gradually as we go farther away from it.

\begin{figure}
\plotone{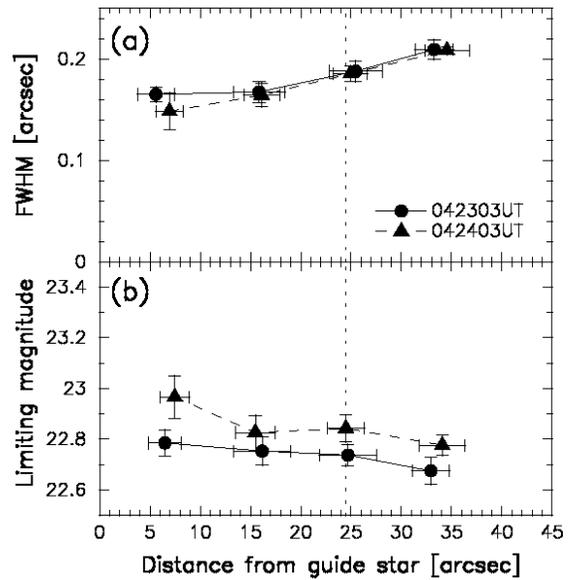}
 \caption{Variation of (a) the FWHM of point sources (spatial
resolution) and (b) the $5\sigma$ limiting magnitude with an hour
integration. The results as a function of distance from a guide star
were based on the snapshots of a globular cluster M13 which were taken
just after the deep imaging on 2003 April 23 (filled circles) and 24
(filled triangles) in Universal Time. Vertical dotted line shows the
location of the reference point-like source (``S'' in Figure
\ref{ssdf_image}), from which the FWHM and $5\sigma$ limiting magnitude
of the SSDF were derived (see \S 3). Most of detected galaxies in the
SSDF are distributed in a range of distance from 10\arcsec\ to
35\arcsec, and the effect of anisoplanatism should be negligible in this
range. \label{aniso}}
\end{figure}

The angular anisoplanatism in our imaging data was evaluated from some
snapshots of globular cluster M13 using AO, in which a number of stars
are distributed uniformly in our 1\arcmin\ $\times$ 1\arcmin\ field of
view. The snapshots were taken just after the deep imaging so that the
observational condition does not change significantly (Table
\ref{log}). The AO guide star for the snapshots was selected to have the
same brightness as that of the deep imaging. Figure \ref{aniso}a shows
the variation of the stellar FWHM (spatial resolution) with respect to
the distance from the guide star. Similarly, Figure \ref{aniso}b shows
the variation of the 5$\sigma$ limiting magnitude with an hour
integration (sensitivity) in the field. The stellar FWHM and the
limiting magnitude deteriorate with increasing distance from the guide
star.

The stellar FWHM and the 5$\sigma$ limiting magnitude in the SSDF are
0\farcs18 and $K^{\prime} \sim 22.9$ with an hour integration,
respectively, for the point-like source located at the distance of
24\arcsec\ from the guide star (dotted line in Figure
\ref{aniso}). Since more than 90\% of the detected galaxies in the SSDF
are distributed in a range of distance from 10\arcsec\ to 35\arcsec, the
uncertainties in measurements of the stellar FWHM and the 5$\sigma$
limiting magnitude due to anisoplanatism are about $\pm$0\farcs03 and
$\pm$0.1 mag, respectively, which should be negligible for the following
discussions.

\section{Discussion}

\subsection{Galaxy number counts}
\begin{figure}
\plotone{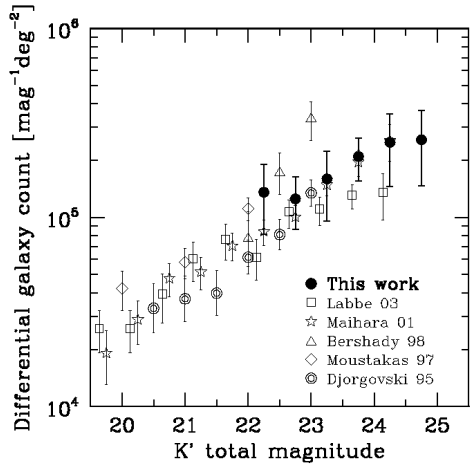}
\caption{Differential $K^{\prime}$-band number counts of galaxies estimated
 from the SSDF galaxy counts in the reliable magnitude range (filled
 circles) where the signal-to-noise ratio for raw counts is more than 3
 and the detection completeness for point source is more than
 50\%. Error bars for this work show the uncertainties due to a poisson
 statistics and a clustering of galaxies estimated from an angular
 correlation function. For a purpose of comparison, the counts in the
 literatures are also plotted. Their magnitude ranges are limited by
 the same reliability criteria as ours.  \label{numcnt}}
\end{figure}

Figure \ref{numcnt} shows our differential number counts of galaxies in
the reliable magnitude range of $22 < K^{\prime} < 25$, where the
signal-to-noise ratio for raw counts is greater than 3 and the detection
completeness for point source is more than 50\%. The
count slope $\alpha \equiv d\log N/dm$ is about 0.15 $\pm$ 0.08, where
the uncertainty was defined as the uncertainty in the least-square
fitting to the data. For a comparison, other $K$ and $K^{\prime}$ band
number counts in the literatures are also shown in Figure \ref{numcnt}
and summarized in Table \ref{comp_slope}. In this paper, we made no
distinction between $K$ and $K^{\prime}$\, because the magnitude difference is
expected to be $\langle K^{\prime}-K \rangle\la 0.03$ \citep{min98a},
which is negligible. 

\begin{deluxetable*}{lcccccc}
\tabletypesize{\scriptsize}
\tablecaption{The comparison of the SSDF with other deep surveys in the
 $K$ band. \label{comp_slope}}
\tablewidth{0pt}
\tablehead{
\colhead{Survey}  & \colhead{Instrument} & \colhead{Integration time} &
 \colhead{Area} & \colhead{$m_{50}$\tablenotemark{a}} &
 \colhead{$\alpha$\tablenotemark{b}} & \colhead{mag
 range\tablenotemark{c}}\\
\colhead{}  & \colhead{} & \colhead{[hours]} & \colhead{[arcmin$^2$]} &
 \colhead{} & \colhead{} & \colhead{}
}
\startdata
This work (SSDF)     & Subaru/IRCS+AO & 26.8 & 1.00 & 25.0 & 0.15  & 22$-$25\\
\citealt{lab03} (FIRES)      & VLT/ISAAC   & 35.6 & 6.25 & 24.4 & 0.15\tablenotemark{d} & 22$-$24 \\
\citealt{mai01} (SDF)    & Subaru/CISCO   & 9.7  & 4.00 & 24.4 & 0.23 & 20$-$24 \\
\citealt{ber98}   & Keck/NIRC    & 4.9  & 1.50 & 24.0 & 0.36 & 19.5$-$22.5 \\
\citealt{mou97}  & Keck/NIRC     & 1.7 & 2.41 & 22.7 & 0.23 & 18$-$23 \\
\citealt{djr95} & Keck/NIRC    & 5.6  & 0.44 & 23.5 & 0.32 & 20$-$24 \\
\enddata
\tablenotetext{a}{Magnitude with 50\% detection completeness for point source.}
\tablenotetext{b}{The slope of number counts $\alpha \equiv d\log N/dm$.}
\tablenotetext{c}{Magnitude range for $\alpha$.}
\tablenotetext{d}{Their slope is 0.25 at brighter magnitude of
 20.0$<K_s<$22.0 and declines to 0.15 at fainter magnitude range of 22.0$<K_s<$24.0.}
\end{deluxetable*}

Our counts at $23 < K^{\prime} < 24.5$ are in good agreement with those
of the SDF \citep{mai01}, although these counts at $22 < K^{\prime} <
23$ differ by a factor of 1.2$-$1.6. Earlier results of the $K$-band
deep imaging by the Keck telescope reported the slope of 0.23
\citep{mou97}, 0.32 \citep{djr95}, and 0.36 \citep{ber98} at $20 <
K^{\prime} < 23$, respectively. The discrepancy of the faint-end slopes
could be originated from the difference in the adopted technique of
incompleteness correction.  The HDF-S deep imaging
\citep[FIRES,][]{lab03} shows that the count slope changes from
$\alpha\,\simeq\,0.25$ at $20 < K^{\prime} < 22$ to
$\alpha\,\simeq\,0.15$ at $22 < K^{\prime} < 24$. The break of the count
slope at the faint end had not been seen in other deep $K$-band imaging
surveys. Although our counts are larger by a factor of about 1.5 than
those of the HDF-S, our count slope at $22 < K^{\prime} < 25$ is
consistent with the HDF-S slope at $22 < K^{\prime} < 24$, supporting
the flatter slope of $\alpha\,\simeq\,0.15$ at $K^{\prime} > 22$. The
break of the count slope at the faint end have also been seen in the
$H$-band galaxy counts in the northern HDF (HDF-N) \citep{tho03} that
extend to even fainter magnitude than ours. Assuming the average $H-K$
color of the galaxies equals to 1, the $H$-band counts in the HDF-N show
the break at the same magnitude as the $K$-band counts.

Some theoretical models predicted further increase of number counts beyond
$K \sim 23$. For example, \citet{bab96} claimed that a large number of
faint blue dwarf galaxies forming at high redshift should become
observable at this $K$-magnitude. Moreover, \citet{tom95} claimed that
the cosmic density at high redshift should be larger than that in the
nearby universe, suggesting an excessive population of distant galaxies
in the faint end. However, our number count slope is as flat as $\alpha
\sim 0.15$ down to $K^{\prime} \sim 25$, suggesting that such extreme
scenarios of high-redshift galaxy formation are unlikely.

\subsection{Extragalactic background light}
Extragalactic background light (EBL) in the optical and NIR wavelengths
is an indicator of the total luminosity in the universe, which is
believed to be dominated by the integration of all stellar light
\citep{bon86,yos88}. Thereby, it provides a quantitative estimate of the
baryonic mass that is a fundamental quantity for galaxy formation and
cosmology. If all stellar light is emitted from galactic systems, the
EBL can be resolved into discrete galaxies by deep imaging survey. The
faint-end slope $\alpha$ of the HDF galaxy counts \citep{wil96} is
flatter than the critical slope of 0.4, with which the contributed flux
from galaxies to the EBL is constant against magnitude. Therefore, the
light from the faint-end galaxies does not significantly increase the
EBL, suggesting that the galaxies that contribute to the EBL have
already been resolved into discrete galaxies \citep{mad00}.  Although
there is a considerable scatter in the faint-end counts, the situation
is the same for the NIR bands. However, the measurements of diffuse EBL
in the optical and NIR bands suggest that the diffuse EBL flux is
consistently higher than the integrated flux of galaxies. For instance,
the $K^{\prime}$-band integrated flux of galaxies was estimated at 7$-$8
nWm$^{-2}$sr$^{-1}$ \citep{mad00,mai01}, while the $K$-band diffuse EBL
was estimated at 20$-$29 nWm$^{-2}$sr$^{-1}$
\citep{gor00,wri01,mat01,cam01}. This discrepancy could be caused by
various selection effects or incompleteness of source
detection. \citet{tot01b} theoretically estimated the number of missed
galaxies in the SDF \citep{mai01} due to selection effects or
incompleteness of the detection of galaxies. They reported that the
$K$-band EBL from the detected galaxies is estimated at 7.8$-$10.2
nWm$^{-2}$sr$^{-1}$ and concluded that more than 90\% of the galaxies
contributed to the EBL has already been resolved into discrete galaxies
and the flux from missed galaxies cannot reconcile the discrepancy
between the integrated flux of galaxies and the diffuse EBL. However,
since there could be some inevitable uncertainties in theories, even
deeper observation is essential to reveal the origin of this
discrepancy.

\begin{figure}
\plotone{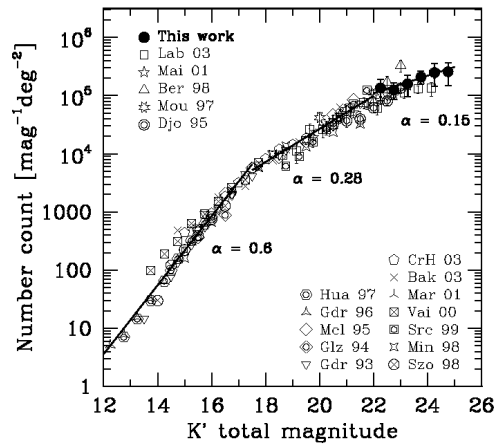}
\caption{Three power-law curves were fitted to the differential number
counts of galaxies at $12 < K^{\prime} < 25$ based on the combined data
of ours and those in the literatures. The count slope $\alpha = d\log
N/dm$ is $0.59 \pm 0.01$ at $K^{\prime} < 17.5$, $0.29 \pm 0.01$ at
$17.5 < K^{\prime} < 22$, and $0.15 \pm 0.08$ at $22 < K^{\prime} <
25$. \label{numcnt_fit}}
\end{figure}

\begin{figure}
\plotone{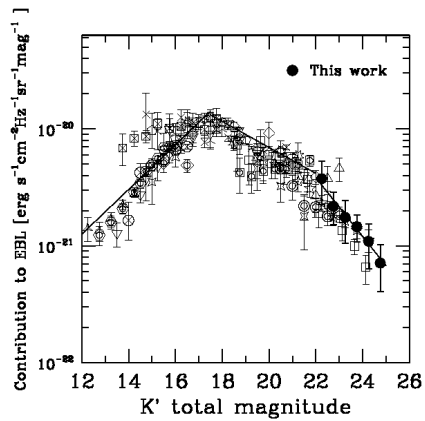}
\caption{Contributed flux of galaxies in each $K^{\prime}$-magnitude bin to the
 extragalactic background light (EBL). The data in the literatures are
 also shown using the same symbols as in Figure \ref{numcnt_fit}.  Solid
 lines correspond to the fitted curves in Figure
 \ref{numcnt_fit}. \label{ebl}}
\end{figure}

In this context, we derived the integrated flux of galaxies contributed
to the EBL from our $K^{\prime}$-band number counts, which is the deepest and
less affected by the incompleteness of the source detection. Figure
\ref{numcnt_fit} shows the count slopes in the separate magnitude
ranges. At brighter magnitude, we used the slope of $\alpha=0.59 \pm
0.01$ at $K^{\prime}<17.5$ and $\alpha=0.29 \pm 0.01$ at
$17.5<K^{\prime}<22$, which were derived from the previous data with the
signal-to-noise ratio of more than 3 and with the detection completeness
for point source of more than 50\%. At fainter magnitude of
$22<K^{\prime}<25$, we used our result of $\alpha=0.15 \pm 0.08$, which
is most reliable among the existing faint-end data.  Figure \ref{ebl}
shows the contributed flux of galaxies in each $K^{\prime}$-magnitude bin to
the EBL. The total contribution was calculated to be 9.26 $\pm$ 0.14
nWm$^{-2}$sr$^{-1}$ using the slopes in the above three magnitude
ranges, where the uncertainty in the flux was determined from the
uncertainty in the slope at the faint end.  Even flatter count slope
than the previous results at the faint end suggests that the contributed
flux of the faint-end galaxies to the EBL is less significant. If the
count slope does not change at $K^{\prime}>25$, the total contribution
is expected to be about 9.41 nWm$^{-2}$sr$^{-1}$, suggesting that more
than 98\% of the galaxies contributed to the EBL has already been
resolved as discrete galaxies in our data at $K^{\prime}<25$. 

Based on our SSDF counts down to $K^{\prime} \sim 25$, which is 0.5 mag
deeper than the SDF, we found that our estimated $K^{\prime}$-band EBL
from galaxies is almost consistent with the theoretical estimation of
\citet{tot01b}. We also found that our estimated $K^{\prime}$-band EBL
from galaxies is close to the $H$-band EBL from galaxies estimated from
the HDF-N imaging (6$-$7 nWm$^{-2}$sr$^{-1}$, \citealt{tho01,tho03}),
which is comparable or even deeper than our $K^{\prime}$-band
imaging. These results support the conclusion of \citet{tot01b} that the EBL
from galaxies has been resolved almost completely into discrete sources
of normal galaxies.  Comparison of our estimated EBL from galaxies with
the measurements of diffuse EBL in the $K$ band (20$-$29
nWm$^{-2}$sr$^{-1}$) shows that a population of galaxies accounts for
only less than 50\% of the diffuse EBL in the $K$ band.  Unless the
diffuse EBL measurements are contaminated by crucial systematic
uncertainties, a population other than known galaxy populations at even
fainter magnitude is necessary to explain such discrepancy. It has been
pointed out that a hypothetical population might consist of population
III stars at very high redshift, provided that a large amount of
baryonic matter comparable to stars in galaxies should be used to form
such exotic stars at extremely high rate \citep{san02,sal03,coo04}.

\subsection{Size distribution}

\begin{figure*}
\plottwo{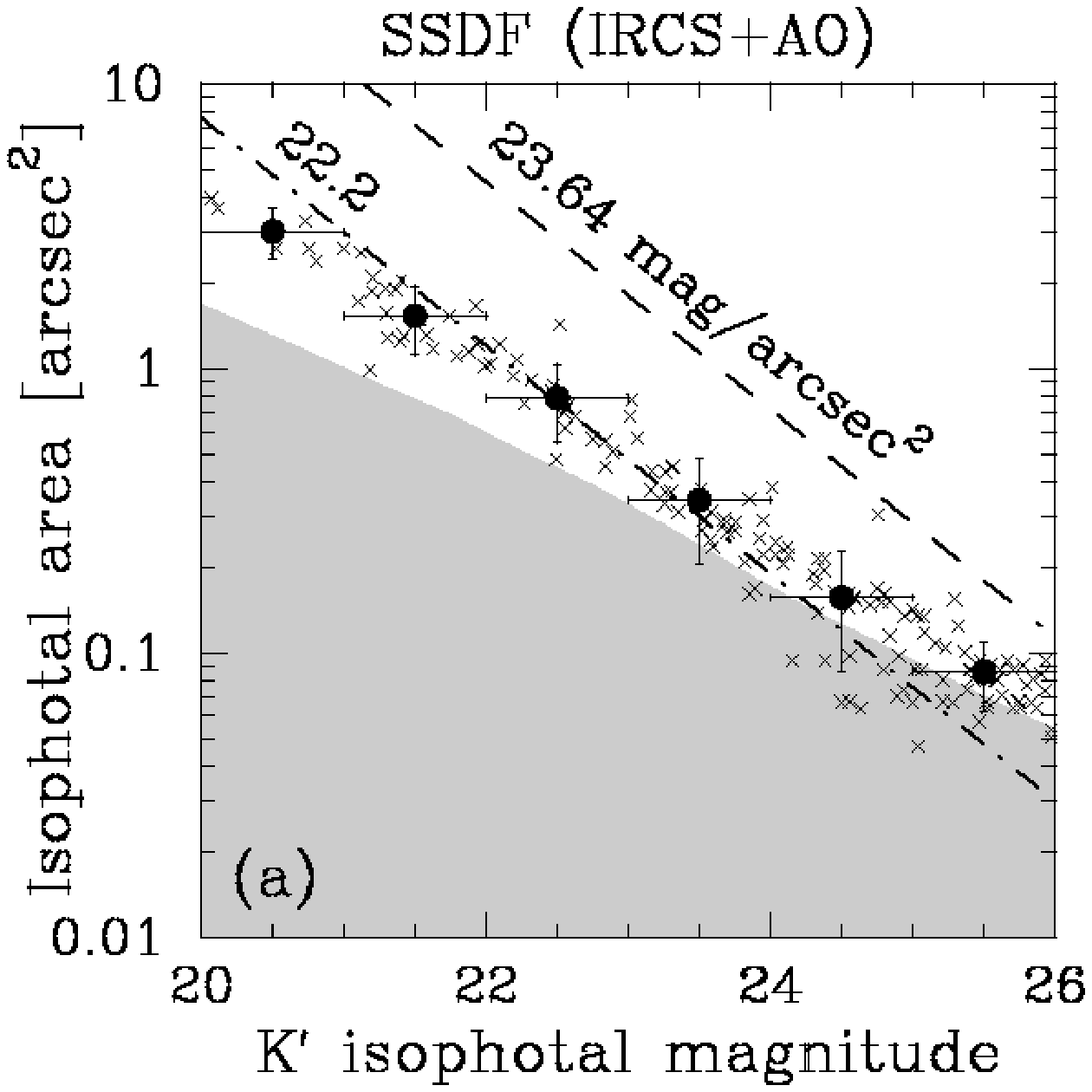}{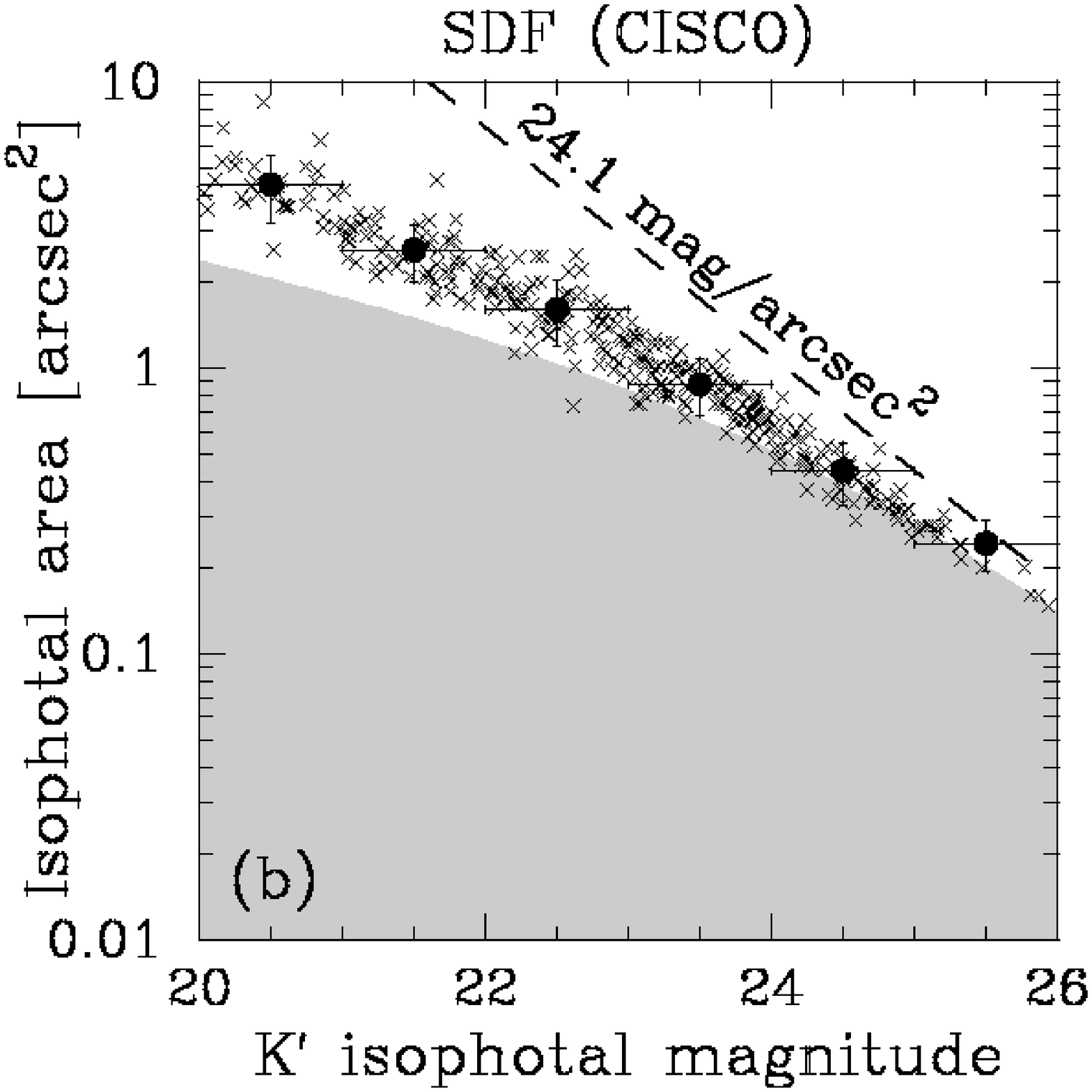}
\caption{ Size distribution of the detected galaxies in the isophotal
area versus isophotal magnitude diagram for (a) the SSDF and (b) the
SDF. The SDF data were taken from \citet{tot01a}. Crosses show the
individual galaxies. Filled circles show the mean size in each one
magnitude bin. The horizontal error bars show the bin size around the
mean and the vertical error bars show the 1$\sigma$ dispersion of the
isophotal area of the detected galaxies. The gray shaded region is where
the isophotal area is less than the expected area of point source for
each isophotal magnitude in the SSDF or the SDF. The dashed line shows
the maximum isophotal area for each isophotal magnitude with the surface
brightness (SB) threshold (23.64 mag/arcsec$^2$ for the SSDF, 24.1
mag/arcsec$^2$ for the SDF).  Thanks to the high spatial resolution due
to AO, we can resolve the galaxies having the area size of less than 0.1
arcsec$^2$ scale, which could not be done with the SDF image. The
dot-dashed line in (a) shows the constant SB of 22.2 mag/arcsec$^2$. The
constant SB line matches the observed size distribution within the
uncertainty, although the observed size of the faint-end galaxies might
be biased by the PSF. This suggests that the SB is not much different
between high-redshift (faint) and low-redshift (bright)
galaxies. \label{sizedist}}
\end{figure*}%

Figure \ref{sizedist} shows the size distribution of detected galaxies
in the isophotal $K^{\prime}$-magnitude ($m_{iso}$) versus isophotal area
($A_{iso}$) diagram for the SSDF and the SDF. The observed mean size is
shown by filled circles. The sizes of individual SDF galaxies (crosses
in Figure \ref{sizedist}b) are highly biased due to the large PSF size
especially at faint magnitude of $K^{\prime}_{iso} > 24$. However,
because such bias is very small for the SSDF data thanks to the high
spatial resolution with AO, the size of galaxies can be distinguished
from that of PSF even in the faint end. The slope of $m_{iso}-A_{iso}$
relation in the SSDF seems to be constant at $20 < K^{\prime}_{iso} <
26$, yielding $d \log A_{iso}/dm_{iso} = -0.32 \pm 0.12$. This slope is
almost consistent with that of constant surface brightness ($d \log
A_{iso}/dm_{iso} = -0.4$) within the uncertainties, while the slope in
the SDF seems to be flatter in the faint end due to the large bias of
the PSF. We note that the constant surface brightness of 22.2
mag/arcsec$^2$ (dot-dashed line in Figure \ref{sizedist}a) matches the
observed size distribution with the confidence level of 95\%. This
implies that the surface brightness is not much different between
high-redshift (faint) and low-redshift (bright) galaxies.

\begin{figure}
\plotone{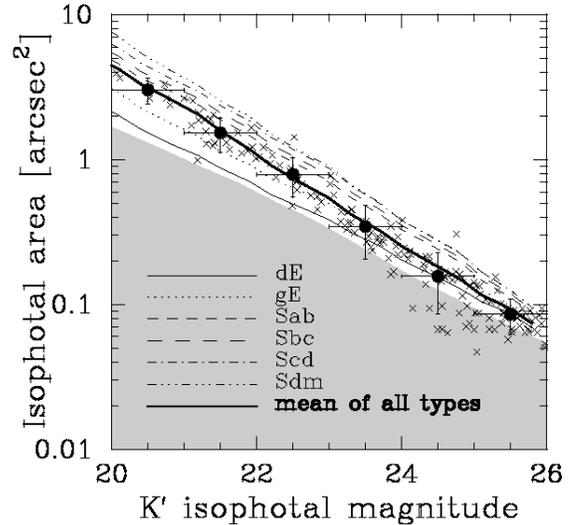}
\caption{ Same figure in Figure \ref{sizedist}(a), but with prediction
 of the PLE model with no size evolution described in the text. Thin
 lines show the predicted mean size for six morphological types of
 galaxies (dE, gE, Sab, Sbc, Scd, Sdm). The thick solid line shows the
 predicted mean size for all types of galaxies. \label{size_type}}
\end{figure}

Figure \ref{size_type} compares the observed size distribution of
galaxies to theoretical predictions from a standard model of pure
luminosity evolution (PLE) of galaxies that assumes no number evolution
or no merging of galaxies. We used the same model as \citet{ty00} and
\citet{tot01a}, which is an updated version of \citet{yos93}. In this
model, galaxies are divided into six morphological types (dE, gE, Sab,
Sbc, Scd, Sdm) and their type-dependent luminosity evolutions are
described by the model of \citet{ari87} and \citet{ari92} in which the
star formation history is determined to reproduce the present-day colors
and chemical properties of galaxies. All galaxies are simply assumed to
be formed at a single redshift of $z_F$=3. The surface brightness
profiles of elliptical and spiral galaxies are modeled by the de
Vaucouleurs and exponential profiles, respectively. These profiles are
convolved with a moffat profile which nicely represents the observed PSF
(see \S6.2), and are used to calculate the isophotal area. The effective
radius of galaxies is determined from the observed luminosity ($L$)
versus effective radius ($r_e$) relation of local galaxies
\citep{ben92,imp96}. Detailed descriptions of the model are given in
\citet{yos93}, \citet{ty00} and \citet{tot01a}. The predicted
$m_{iso}-A_{iso}$ relations for different types of galaxies are shown in
Figure \ref{size_type}. We found that the predicted mean size for all
types (thick solid line) is in good agreement with the observed
mean size (filled circles).

\begin{figure}
\plotone{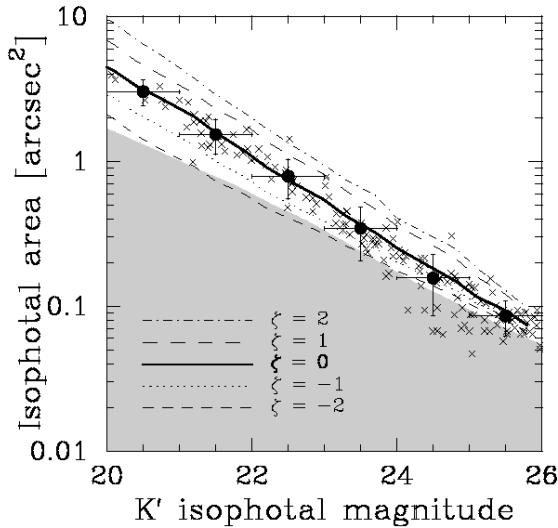}
\caption{ Same figure as Figure \ref{sizedist}(a), but with prediction
of the PLE model which allows for size evolution, assuming that the
effective radius of galaxies, $r_e$, evolves as $r_e \propto
(1+z)^{\zeta}$. Solid line, which is identical to the thick solid line
in Figure \ref{size_type}, shows the predicted mean size for all types
of galaxies assuming no intrinsic size evolution ($\zeta = 0$). The
long-dashed, dot-dashed, dotted, and short-dashed lines show the models
with size evolution of $\zeta$=1, 2, -1, and -2,
respectively. \label{size_evol}}
\end{figure}

Figure \ref{size_evol} compares the observed size distribution to a PLE
model that allows for intrinsic size evolution, in which galaxy size
changes without changing the total luminosity (i.e. a size evolution is
not driven by a number evolution).  We here introduce a phenomenological
parameter $\zeta$, and multiply an additional factor $(1+z)^{\zeta}$ to
the size of PLE model galaxies used in Figure \ref{size_type}. It is
seen from Figure \ref{size_evol} that the observed size distribution
agrees with the model of $\zeta=0$, which favors no intrinsic size
evolution, in agreement with the earlier claim by \citet{yos93}. This
suggests that the $L-r_e$ relation of high-redshift galaxies is almost
same as the local relation. Therefore, the result in Figure
\ref{sizedist}a that the surface brightness is not much different
between high- and low-redshift galaxies is further strengthened.

\citet{tot01a} performed the same analysis as ours using the SDF data
taken by the Subaru/CISCO. They claimed that the size distribution of
galaxies has the same trend as ours, although it was difficult to
distinguish between the models of different types or different $\zeta$
based on their faint-end data at $K^{\prime}_{iso} \geq 24$, because the
size of detected galaxies in the SDF was almost close to the size of PSF
with FWHM $\sim$ 0\farcs45. For our SSDF data with AO, the PSF size is
small enough to be separated from the galaxies at $K^{\prime}_{iso} \geq
24$, thereby confirming the claim by \citet{tot01a} unambiguously.

We note that a size evolution driven by a number evolution or merging is
not necessarily ruled out if such process proceeds without changing the
magnitude versus size relation in Figure \ref{size_evol}.  In fact, this
relation is maintained if the typical surface brightness of galaxies is
preserved by mergers. What we can conclude from our result is that the
surface brightness of the faintest SSDF galaxies is not much different
from that expected for a simple PLE model without size
evolution. Therefore, our result places a strong constraint on the
contemporary CDM-based structure formation theory (e.g.,
\citealt{whi78,nag02}), in which galaxies are formed through merging of
many smaller building blocks at different formation epochs. Further
information can be obtained by the relation between redshift and
effective radius of galaxies, which should be obtained from our SSDF
data through the galaxy profile fitting (see Figure \ref{radfit}) and
the photometric redshift technique (Minowa et al. 2005, in preparation).

\section{Summary}
We have presented the first result of deep $K^{\prime}$-band imaging of the
SSDF with the Subaru AO. The achieved detection limit is $K^{\prime}
\sim 24.7$ (5$\sigma$, 0\farcs2 aperture) for point source and
$K^{\prime} \sim 23.5$ (5$\sigma$, 0\farcs6 aperture) for galaxies with the total
integration time of 26.8 hours. The sensitivity gain due to the AO
correction is about 0.5 mag. The achieved spatial resolution is
0\farcs18 in stellar FWHM on average. We found from the simulation that
the extended wings of the observed AO PSF does not affect the measurement
of the flux and size of the typical galaxies in the SSDF. Combination of
high sensitivity and high spatial resolution by AO enables detailed
morphological study of high-redshift galaxies with the profile fitting
technique. From a similar AO observation of globular cluster M13, we
confirmed that the variation of the spatial resolution and the
sensitivity to the point source (anisoplanatism) is not significant
within our 1\arcmin $\times$ 1\arcmin\ field of view.

The slope of differential galaxy number counts ($\alpha \equiv d \log
N / dm$) is about 0.15 at $22<K^{\prime}<25$. Our count slope is consistent
with that of the HDF-S \citep[FIRES,][]{lab03}, in which the slope
changes from $\alpha \simeq 0.25$ at $20<K^{\prime}<22$ to $\alpha
\simeq 0.15$ at $22<K^{\prime}<24$.  Thus, our data support the flatter
slope of $\alpha \simeq 0.15$ at $K^{\prime} > 22$ with the deepest
image to date, while the other $K$-band deep surveys show the steeper
slope at the faint-end. The discrepancies of faint-end counts between
ours and other surveys might be caused by either field-to-field
variation or different technique of incompleteness correction.

The contributed flux of galaxies to the EBL is about 9.43
nWm$^{-2}$sr$^{-1}$, and 98\% of which appears to be resolved into
discrete galaxies in the SSDF at $K^{\prime}<25$. However, our EBL
estimate can account for only less than 50\% of the total flux of
diffuse EBL which was independently measured by satellites in the
similar $K$ band. Our result further strengthens the previous claim by
\citet{tot01b} that there must be an exotic population that
significantly contributes to the EBL, unless the measurements of diffuse
EBL are contaminated by crucial systematic uncertainties.

We examined the size distribution of galaxies in the isophotal magnitude
versus isophotal area diagram. In this diagram, the size distribution of
detected galaxies was obtained down to the area size of less than 0.1
arcsec$^2$, which is less than a half of that of the previous SDF
galaxies. We found that the relation of observed mean size as a function
of magnitude is nicely explained by a simple PLE model. We examined a
possibility of intrinsic size evolution using a PLE model that allows
for size evolution, and found that the model with no size evolution
gives the best fit to the data. We conclude that the surface brightness
of galaxies at high redshift is not much different from that expected
from the size-luminosity relation of present-day galaxies. We cannot
rule out a size evolution associated with a number evolution of galaxies
if their surface brightness is preserved. In other words, our result of
constant surface brightness should place a strong constraint on the
merging process in a hierarchical structure formation theory.

\acknowledgments

 We would like to thank the Subaru time allocation committee for their
judgment to approve our proposal of the SSDF project as an intensive
program (proposal ID : S02A-IP1, S03A-062, PI : Yoshii, Y.). We are
indebted to H. Karoji, Director of the Subaru Telescope, for giving us
the time for test observations of the SSDF project. Although we did not
use such test data in this paper, it significantly helped us to make a
success of this project. We appreciate the Subaru AO/IRCS team members
and the Subaru staffs for their kind assistance during the observations
and data reduction. This project would not have been possible without
their dedicated support. This project was financially supported by COE
Research (07CE2002) of the Ministry of Education, Science, and Culture
of Japan. Y. M. is financially supported by the Japan Society for the
Promotion of Science (JSPS).

\clearpage


\begin{thebibliography}{}
\bibitem[Arimoto \& Yoshii(1987)]{ari87} Arimoto, N., \& Yoshii,
				 Y. 1987, \aap, 173, 23 
\bibitem[Arimoto et al.(1992)]{ari92} Arimoto, N., Yoshii, Y., \&
				 Takahara, F. 1992, \aap, 253, 21
\bibitem[Babul \& Ferguson(1996)]{bab96} Babul, A. \& Ferguson, H. C.\
				 1996, \apj, 458, 100
\bibitem[Baker et al.(2003)]{bak03} Baker, A. J. et al. 2003, \aap, 406,
				 593
\bibitem[Baugh et al.(1996)]{bau96} Baugh, C. M., Gardner, J. P., Frenk,
				 C. S., \& Sharples, R. M. 1996, \mnras,
				 283, L15
\bibitem[Bernstein et al.(2002)]{ber02} Bernstein, R. A., Freedman,
				 W. L., \& Madore, B. F. 2002, \apj,
				 571, 107 
\bibitem[Bershady et al.(1998)]{ber98} Bershady, M. A., Lowenthal,
				 J. D., \& Koo, D. C. 1998, \apj, 505,
				 50
\bibitem[Bender et al.(1992)]{ben92} Bender, R., Burstein, D., \& Faber,
				 S. M. 1992, \apj, 399, 462 
\bibitem[Bertin \& Arnouts(1996)]{ber96} Bertin, E \& Arnouts, S 1996,
				 \aaps, 117, 393
\bibitem[Bond et al.(1986)]{bon86} Bond, J. R., Carr, B. J., \& Hogan,
				 C. J.\ 1986, \apj, 306, 428 
\bibitem[Cambr{\' e}sy et al.(2001)]{cam01} Cambr{\' e}sy, L., Reach,
				 W. T., Beichman, C. A., \& Jarrett,
				 T. H. 2001, \apj, 555, 563
\bibitem[Carlberg et al.(1997)]{car97} Carlberg, R. G., Cowie, L. L.,
				 Songaila, A., \& Hu, E. M. 1997, \apj, 484, 538 
\bibitem[Cooray \& Yoshida(2004)]{coo04} Cooray, A. \& Yoshida, N. 2004,
				 \mnras, 351, 71
\bibitem[Crist{\' o}bal-Hornillos et al.(2003)]{crh03} Crist{\'
				 o}bal-Hornillos, D., Balcells, M.,
				 Prieto, M., Guzm{\' a}n, R., Gallego,
				 J., Cardiel, N., Serrano, {\' A}., \&
				 Pell{\' o}, R.\ 2003, \apj, 595, 71
\bibitem[de Vaucouleurs(1948)]{dev48} de Vaucouleurs, G. 1948, Annales
				 d'Astrophysique, 11, 247
\bibitem[Djorgovski et al.(1995)]{djr95} Djorgovski, S. et al. 1995
				 \apj, 438, 13
\bibitem[Freeman(1970)]{fre70} Freeman, K. C. 1970, \apj, 160, 811
\bibitem[Fugal \& Moody(2003)]{fug03} Fugal, J. P., \& Moody,
				 J. W. 2003, \pasp, 115, 295
\bibitem[Gardner et al.(1993)]{gdr93} Gardner, J. P., Cowie, L. L., \&
				 Wainscoat, R. J. 1993, \apj, 415, L9
\bibitem[Gardner et al.(1996)]{gdr96} Gardner, J. P., Sharples, R. M.,
				 Carrasco, B. E., \& Frenk, C. S. 1996,
				 \mnras, 282, L1
\bibitem[Gardner et al.(2000)]{gdr00} Gardner, J. P., et al. 2000, \aj,
				 119, 486
\bibitem[Glazebrook et al.(1994)]{glz94} Glazebrook, K., Peacock, J. A.,
				 Miller, L., \& Collins, C. A. 1994,
				 \mnras, 266, 65
\bibitem[Gorjian et al.(2000)]{gor00} Gorjian, V., Wright, E. L., Chary,
				 R. R. 2000, \apj, 536, 550
\bibitem[Groth \& Peebles(1977)]{gro77} Groth E. J., Peebles
				 P. J. E. 1977, \apj 217, 38
\bibitem[Hardy(1998)]{har98} Hardy, J. W. 1998, Adaptive Optics for
				 Astronomical Telescopes (New York:
				 Oxford University Press)
\bibitem[Huang et al.(1997)]{hua97} Huang, J. S., Cowie, L. L., Gardner,
				 J. P., Hu, E. M., Songaila, A., \&
				 Wainscoat, R. J. 1997, \apj, 476, 12
\bibitem[Impey et al.(1996)]{imp96} Impey, C. D., Sprayberry, D., Irwin,
				 M. J., \& Bothun, G. D. 1996, \apjs,
				 105, 209 
\bibitem[Iye et al.(2004)]{iye04} Iye, M. et al. 2004, \pasj, 56, 381
\bibitem[Kahikawa et al.(2004)]{kas04} Kashikawa, N. et al. 2004, \pasj, 56, 1011
\bibitem[Kajisawa \& Yamada(2001)]{kaj01} Kajisawa, M. \& Yamada, T. \pasj,
				 53, 833
\bibitem[Kobayashi et al.(2000)]{kob00} Kobayashi, N. et al. 2000 \procspie, 4008, 1506
\bibitem[Labbe et al.(2003)]{lab03} Labbe, I. et al. 2003 \aj, 125, 1107
\bibitem[Madau \& Pozzetti(2000)]{mad00} Madau, P. \& Pozzetti, L. 2000,
				 \mnras, 312, L9
\bibitem[Maihara et al.(2001)]{mai01} Maihara, T. et al. 2001 \pasj
				 53, 25
\bibitem[Martini(2001)]{mar01} Martini, P. 2001, \aj, 121, 598
\bibitem[Matsumoto et al.(2001)]{mat01} Matsumoto, T., et al. 2001, in
				 ISO Surveys of a Dusty Universe,
				 ed. D. Lemke, M. Stickel, \& K. Wilke
				 (New York: Springer)
\bibitem[Mcleod et al.(1995)]{mcl95} McLeod, B. A., Bernstein, G. M.,
				 Rieke, M. J., Tollestrup, E. V., \&
				 Fazio, G. G. 1995, \apjs, 96, 117
\bibitem[Minezaki et al.(1998a)]{min98a} Minezaki, T., Kobayashi, Y.,
				 Yoshii Y., \& Peterson B. A. 1998 \apj,
				 494, 111
\bibitem[Minezaki et al.(1998b)]{min98b} Minezaki, T., Yoshii, Y.,
				 Cohen, M., Kobayashi, Y., Peterson,
				 B. A. 1998, \aj, 115, 229
\bibitem[Miyazaki et al.(2002)]{miy02} Miyazaki, S. et al. 2002 \pasj,
				 54, 833
\bibitem[Motohara et al.(2002)]{mot02} Motohara, K. et al. 2002 \pasj,
				 54, 315
\bibitem[Moustakas et al.(1997)]{mou97} Moustakas, L. A., Davis, M.,
				 Graham, J. R., Silk, J., Peterson,
				 B. A., \& Yoshii, Y. 1997, \apj, 475,
				 445
\bibitem[Nagashima et al.(2002)]{nag02} Nagashima, M., Yoshii, Y.,
				 Totani, T., \& Gouda, N. 2002, \apj,
				 578, 675
\bibitem[Persson et al.(1998)]{per98} Persson S. E., Murphy, D. C.,
				 Krzeminski, W., Roth, M., \& Rieke,
				 M. J. 1998, \aj, 116, 2475
\bibitem[Roche \& Eales(1999)]{roc99a} Roche, N. \& Eales, S. A. 1999, \mnras, 307, 703 
\bibitem[Roche et al.(1999)]{roc99b} Roche, N., Eales, S. A., Hippelein,
				 H., \& Willott, C. J. 1999, \mnras, 306, 538  
\bibitem[Salvaterra \& Ferrara.(2003)]{sal03} Salvaterra, R. \& Ferrara,
				 A. 2003, \mnras, 339, 973
\bibitem[Santos et al.(2002)]{san02} Santos, M. R., Bromm, V., \&
				 Kamionkowski, M. 2002, \mnras, 336, 1082 
\bibitem[Saracco et al.(1999)]{src99} Saracco, P., D'Odorico, S.,
				 Moorwood, A., Buzzoni, A., Cuby, J.-G.,
				 \& Lidman, C. 1999, \aap, 349
\bibitem[Sheth et al.(2003)]{she03} Sheth, K., Regan, M. W., Scoville,
				 N. Z., \& Strubbe, L. E.\ 2003, \apjl,
				 592, L13
\bibitem[Smail et al.(1995)]{sma95} Smail, I., Hogg, D. W., Yan, L., \&
				 Cohen, J. G. 1995, \apj, 449, 105
\bibitem[Szokoly et al.(1998)]{szo98} Szokoly, G., Subbarao, M.,
				 Connoly, A., \& Mobasher, B. 1998,
				 \apj, 492, 452
\bibitem[Takami et al.(2004)]{tak04} Takami, H. et al. 2004 \pasj, 56,
				 225
\bibitem[Thompson(2003)]{tho03} Thompson, R. I. 2003 \apj, 596, 748
\bibitem[Thompson et al.(2001)]{tho01} Thompson, R. I., Weymann, R. J.,
				 \& Storrie-Lombardi, L. J. 2001, \apj,
				 546, 694
\bibitem[Tokunaga et al.(1998)]{tok98} Tokunaga, A. T. et al. 1998
				 \procspie, 3354, 512
\bibitem[Tokunaga et al.(2002)]{tok02} Tokunaga, A. T., Simons, D. A.,
				 \& Vacca, W. D. 2002, \pasp, 114, 180
\bibitem[Tomita(1995)]{tom95} Tomita, K.\ 1995, \apj, 451, 1 
\bibitem[Totani \& Yoshii(2000)]{ty00} Totani, T. and Yoshii, Y. 2000,
				 \apj, 540, 81
\bibitem[Totani et al.(2001a)]{tot01a} Totani, T., Yoshii, Y., Maihara,
				  T., Iwamuro, F., \& Motohara, K. 2001
				  \apj, 559, 592
\bibitem[Totani et al.(2001b)]{tot01b} Totani, T., Yoshii, Y., Maihara,
				  T., Iwamuro, F., \& Motohara, K. 2001
				  \apj, 550, 137
\bibitem[V\"{a}is\"{a}nen et al.(2000)]{vai00} V\"{a}is\"{a}nen, P.,
				 Tollestrup, E. V., Willner, S. P., \&
				 Cohen, M. 2000, \apj, 540, 593
\bibitem[White \& Rees(1978)]{whi78} White, S. D. M. \& Rees,
				 M. J. 1978, \mnras, 183, 341
\bibitem[Williams et al.(1996)]{wil96} Williams, R. T., et al. 1996,
				 \aj, 112, 1335
\bibitem[Williams et al.(2000)]{wil00} Williams, R. T., et al. 2000,
				 \aj, 120, 2735
\bibitem[Wright(2001)]{wri01} Wright, E. L. 2001, \apj, 553, 538
\bibitem[Yan et al.(1998)]{yan98} Yan, L., McCarthy, P. J.,
				 Storrie-Lombardi, L. J., \& Weymann,
				 R. J. 1998, \apj, 503, 19
\bibitem[Yoshii \& Takahara(1988)]{yos88} Yoshii, Y. \& Takahara, F. 1988,
				 \apj, 326, 1
\bibitem[Yoshii(1993)]{yos93} Yoshii, Y. 1993, \apj, 403, 552
\end{thebibliography}
\end{document}